\documentclass[%
 reprint,
 showkeys,
aps,
pra,
floatfix,
]{revtex4-2}

\usepackage[utf8]{inputenc}
\usepackage[margin=1in]{geometry}
\usepackage{blindtext}
\usepackage[utf8]{inputenc}
\usepackage[utf8]{inputenc}
\usepackage{braket}
\usepackage{amsmath} 
\usepackage{graphicx}
\usepackage{float}
\usepackage{breqn}
\usepackage{physics}
\usepackage{xcolor}
\usepackage{amsmath}
\usepackage{amssymb}
\usepackage{amsthm}
\usepackage{linop}
\usepackage{hyperref}
\usepackage{mathtools} 

\makeatletter
\let\cat@comma@active\@empty
\makeatother

\begin{document}
\renewcommand{\refeq}[1]{Eq.~(\ref{eq:#1})}   

\newcommand{\myhighlight}{}


\title{The Lieb-Robinson correlation function for the quantum transverse field Ising model}  

\author{Brendan J. Mahoney}
\affiliation{ 
Department of Electrical Engineering\\
University of Notre Dame\\
Notre Dame, IN~46556, USA
}%
\author{Craig S. Lent}
\affiliation{ 
Department of Electrical Engineering\\
Department of Physics and Astronomy\\
University of Notre Dame\\
Notre Dame, IN~46556, USA
}%

\date{\today}

\begin{abstract}
The Lieb-Robinson  correlation function is the norm of a commutator between local operators acting on separate subsystems at different times. This provides a useful state-independent measure for characterizing the specifically quantum interaction between spatially separated qubits. The finite propagation velocity for this correlator defines a “light-cone” of quantum influence.  We calculate the Lieb-Robinson  correlation function for one-dimensional qubit arrays described by the transverse field Ising model. Direct calculations of  this correlation function have been limited by the exponential increase in the size of the state space with the number of qubits. We introduce a new technique that avoids this barrier by transforming the calculation to a sum over Pauli walks which results in linear scaling with system size. We can then explore propagation in arrays of hundreds of qubits and observe the effects of the quantum phase transition in the system. 
\begin{myhighlight}
We observe the emergence of two distinct velocities of propagation: a correlation front velocity, which is affected by the phase transition, and the Lieb-Robinson velocity which is not.  The correlation front velocity 
is equal to the maximum group velocity of single quasiparticle excitations. The Lieb-Robinson velocity describes the extreme leading edge of correlations when the value of the correlation function itself is still very small. 
\end{myhighlight} 
For the semi-infinite chain of qubits at the quantum critical point, we derive an analytical result for the correlation function. 
\end{abstract}


\maketitle

\section{Introduction}
How does quantum information or influence spread? A significant advance in understanding this important question was famously provided in 1972 by Lieb and Robinson \cite{LiebRobinson1972}. They focused on  many-body systems composed of localized but interacting sub-systems.  
They considered the norm of the commutator
\begin{equation}
    C_{A,B}(t) = \left\|\left[ \hat{A}_k, \hat{B}_m(t)  \right] \right\|
       \label{eq:LRcommutator}
\end{equation}
where  $\hat{A}_k$ operates on subsystem $k$ and 
\begin{equation}
\hat{B}_m(t)=e^{i\hat{H} t/\hbar} \hat{B}_m(0) e^{-i\hat{H} t/\hbar}
\label{eq:HeisenburgTimeDependence}
\end{equation}
is a Heisenberg operator which acts on  subsystem $m$ at time $t$. They showed that under very general conditions,  the norm of the commutator in \refeq{LRcommutator} is bounded,
\begin{equation}
    C_{A,B}(t) \leq 
       K e^{-\mu (d_{k,m} - v_{\text{\tiny LR}} t)},
       \label{eq:BasicLRresult}
\end{equation}
where $d_{k,m}$ is an appropriate distance between the two subsystems.   The Lieb-Robinson velocity $v_{\text{\tiny LR}}$ is understood to be an effective speed limit on the propagation of quantum information or influence. This bound holds even for non-relativistic quantum mechanics and is often envisioned as a sort of effective ``light-cone'' of possible interaction outside of which influence is exponentially suppressed. 

There is by now a large literature calculating and improving the values of $v_{\text{\tiny LR}} $ in the Lieb-Robinson bound for many different systems and including the effects of finite temperatures and disorder \cite{Wang2020,Lucas2020,Lucas2021,Lucas2021b, Gorshkov2023}. The Lieb-Robinson result has been shown to imply finite-length correlations in the ground state of any gapped many-body system \cite{Hastings2006, Nachtergaele2006}. 
Considerable work has been done on specific lattice spin systems, with near-neighbor coupling and those with power-law interactions \cite{FossFeig2015, LuitzPRR2020}. Work on propagation in such spin models is particularly relevant to current developments in quantum computing and quantum information processing.   Recent reviews of the Lieb-Robinson bound and its connection to fundamental questions of operator locality and quantum information are given in references \cite{Lucas2023} and \cite{Swingle2024}.   The bounds have been explored experimentally \cite{Them2014}; a review is  in \cite{Cheneau2022}. 

The Lieb-Robinson analysis is related to issues of quantum entanglement \cite{HorodeckiRMP2009}. Equations (\ref{eq:LRcommutator}) and (\ref{eq:HeisenburgTimeDependence}) involve  only  $\hat{A}_k$,  $\hat{B}_m$, and the system Hamiltonian $\hat{H}$. Entanglement by contrast is a feature of the quantum \emph{state} of a composite system,  which develops precisely because of the interaction between subsystems. The Lieb-Robinson commutator captures an aspect of the effective, possibly mediated, interaction between the two subsystems which is the precondition for them becoming entangled with one another. As emphasized by Chen, Lucas, and Yin \cite{Lucas2023},  the quantity $C_{A,B}(t)$ represents a fundamental constraint on purely quantum non-locality and entanglement.

Here we focus not on the bound in \refeq{BasicLRresult} in the general case but on one specific model--the quantum transverse field Ising (QTFIM) model for a one-dimensional chain of qubits (or equivalently, spins).
Following the work of Colmenarez and Luitz \cite{LuitzPRR2020}, we define the Lieb-Robinson correlation function by
\begin{equation}
    C_k(t) \equiv \left\|  \left[ 
            \hat{\sigma}^z_1(t) \;,
               \hat{\sigma}^z_k\right] \right\|,
               \label{eq:LRcorrDef}
\end{equation}
where the operators are the usual Pauli operators on qubits $1$ and $k$. 

The QTFIM is the simplest model that includes both intrinsic local dynamics and coupling between subsystems, and it has other advantages as well. There is a well-known quantum phase transition  that occurs when the strength of the local dynamics is equal to the inter-qubit coupling. A disordered system for small coupling becomes ferromagnetically ordered at a critical value of the coupling. The Hamiltonian can be diagonalized using now-standard methods which map the many-spin system onto free fermion quasiparticles \cite{LiebMattis1961, SachdevBook2011}.  Though simple in form, the QTFIM is also experimentally and practically relevant; it is used in D-Wave quantum-annealing based computers.  Recent results have been reported for one-dimensional Ising chains consisting of 2000 qubits \cite{King2022} and higher-dimensional configurations with 5000 qubits \cite{King2023}. Quantum-dot cellular automata dynamics have been mapped to the QTFIM \cite{Tougaw1996}. 

We study the behavior of $C_k(t)$ as a function of space (qubit index $k$) and time. Our goals are  to understand in detail the propagation of these quantum correlations down the chain and in particular  (a)  the effects of varying the Hamiltonian parameters, (b) the fingerprints of the quantum phase transition, and (c) the relation of the propagation speed to the Lieb-Robinson velocity.

Directly evaluating $C_k(t)$ has been practically challenging for all but relatively short chains even for the QTFIM. The reasons are familiar: the size of the state space grows exponentially with the length of the chain, and the matrix exponentials in \refeq{HeisenburgTimeDependence} are costly to compute. Heroic methods may be required for chains of 22 qubits \cite{LuitzPRR2020}. For the QTFIM, we devise an operator Pauli walk method which scales only linearly with the system size, and so allows evaluation of \refeq{LRcorrDef} for chains of hundreds of qubits and long times. This is sufficient to establish key features of the propagation for chains of any length.  For the special case of the QTFIM at the quantum critical point, the method is extended to yield a closed form expression for the semi-infinite chain.

The term \emph{correlation function}  often denotes an expectation value, for example of the form
\begin{equation}
\left<\hat{A}(0) \hat{B}(t)\right>  - \left<\hat{A}(0)\right> \left<\hat{B}(t)\right>,
\end{equation}
and so depends on the system state \cite{Schweigler2017}. By contrast, the Lieb-Robinson correlation function defined in \refeq{LRcorrDef} is a correlation between operators independent of state. Moreover, because of the commutator,  it captures only explicitly quantum mechanical effects not classical correlations. It is also fairly associated with bounding the spread of  \emph{information} in a quantum  channel \cite{ChessaGiovannetti2019}, though it is not defined in Shannon information theory terms. The quantity $C_k(t)$ could also be conceived as quantifying \emph{influence}, or even \emph{potential} influence of one system on another (and reciprocally). We will refer to it as a correlation function; its precise meaning is clear from the definition. 

We consider the dynamics of the QTFIM in the sense of calculating $C_k(t)$ using the time-dependent Heisenberg operators. The Hamiltonian itself is time-independent. We are not addressing the situation of the time-dependent state of the system when the Hamiltonian changes abruptly---a so-called ``quench.''

The organization of the paper is as follows. Section \ref{sec:Model} describes the QTFIM and the corresponding operator space. Section \ref{sec:PauliWalkMethod} presents the operator Pauli walk method which is at the heart of our calculations. The special case of evaluating $C_k(t)$ at the quantum critical point is described in Section \ref{sec:CriticalPointCalculation}. At the leading edge of the correlation front which propagates down the chain, a simpler analysis is possible. This was described in \cite{Mahoney2022} in one, two, and three dimensions. We review the main results for one-dimensional chains in Section \ref{sec:Leading edge} in order to connect them with the present calculations. In Section \ref{sec:Propagation}, we examine  the effect of the inter-qubit coupling strength on $C_k(t)$, its saturation value, and propagation speed. A discussion of the main results follows.

%


\section{Model \label{sec:Model}}
\subsection{System Hamiltonian}
We consider a linear array of $N_q$ qubits (or spins) with near-neighbor interactions. The system is described by the transverse-field Ising model with Hamiltonian 

\begin{equation}
\hat{H} = - \gamma \sum_{k = 1}^{N_q} \op{\sigma}{k}{x} - J \sum_{k= 1}^{N_q -1} \op{\sigma}{k}{z} \op{\sigma}{k+1}{z}.
\label{eq:IsingHamiltonian}
\end{equation}
where the operators $\op{\sigma}{k}{\{x,y,z\}}$ are the usual Pauli operators on site $k$. The first term in Eq.~(\ref{eq:IsingHamiltonian}) represents the internal dynamics of each qubit and produces a characteristic time for these dynamics given by 
\begin{equation}
    \tau \equiv \frac{\pi \hbar}{\gamma}.
    \label{eq:tauDef}
\end{equation}
The second term represents the interaction between adjacent qubits and results in an energetic cost for neighboring opposite spins.  We can re-write Eq.~(\ref{eq:IsingHamiltonian}) in dimensionless form as
\begin{equation}
\hat{H}'= \hat{H}/\gamma= - \sum_{k = 1}^{N_q} \op{\sigma}{k}{x} - J' \sum_{k = 1}^{N_q -1} \op{\sigma}{k}{z} \op{\sigma}{k+1}{z},
\label{eq:IsingHamiltonianPrime}
\end{equation}
with $J'=J/\gamma$ now characterizing the system. The quantum phase transition between the ordered and disordered phase occurs at $J'=1$.



\subsection{Operator space}
The set of operators on the qubit array form a Hilbert space with the inner product 
\begin{equation}
    \Braket{\hat{A} | \hat{B}} \equiv \frac{1}{\cal{N}} \Tr
    \left( \hat{A}^\dagger \hat{B} \right).
    \label{eq:TraceInnerProductDef}
\end{equation}
with ${\mathcal{N}} = 2^{N_q}$ being the dimension of the operator state space. 
The norm induced by this inner product is
\begin{equation}
    \| \hat{A}  \| = \sqrt{\frac{1}{\cal{N}} \Tr
    \left( \hat{A}^{\dagger} \hat{A} \right)} =  \sqrt{\frac{1}{\cal{N}} \Tr
     \left| \hat{A} \right|^2 },
         \label{eq:NormDef}
\end{equation}
known as the normalized Frobenius norm. The operator norm, which is equal to the maximum modulus of the singular values, is frequently used in defining the Lieb-Robinson bound. Different norms can in general produce somewhat different results \cite{Lucas2020}, but for the problem at hand the two norms 
\begin{myhighlight}
    yield identical values (see Section \ref{sec:Norms} below and the Supplementary Material \cite{SupplementalMaterial}).
\end{myhighlight}


Pauli operators on the same site $k$ obey the commutation relations
\begin{align}
\left[\hat{\sigma}^{x}_{k},\hat{\sigma}^{y}_{k} \right]=2i\hat{\sigma}^{z}_{k},
\; \left[\hat{\sigma}^{y}_{k},\hat{\sigma}^{z}_{k} \right]=2i\hat{\sigma}^{x}_{k},
\; \left[\hat{\sigma}^{z}_{k},\hat{\sigma}^{x}_{k} \right]=2i\hat{\sigma}^{y}_{k},
\label{eq:PauliCommRelation}
\end{align}
and Pauli operators on different sites commute. It is sometimes convenient to use  the common alternative notation 
\begin{equation}
 X_k=\hat{\sigma}^{x}_{k},  
 \qquad Y_k=\hat{\sigma}^{y}_{k}, 
 \qquad Z_k=\hat{\sigma}^{z}_{k}.
\end{equation}
We consider Pauli strings of the form
\begin{equation}
\hat{\sigma}_s = \prod_{k=1}^{N_q} \hat{\sigma}^{\alpha_k}_{k}, 
\qquad
\hat{\sigma}^{\alpha_k}_{k} \in [\hat{I}_k,\hat{\sigma}^{x}_k, \hat{\sigma}^{y}_k, \hat{\sigma}^{z}_k ].
\label{eq:pauli_string}
\end{equation}
For example, using the simpler notation,
\begin{equation}
    \hat{\sigma}_s= X_1 X_2 Z_3 I_4 \ldots 
\end{equation}
or
\begin{equation}
    \hat{\sigma}_{s'}= Z_1 I_2 Y_3 Z_4 \ldots .
\end{equation}
The set of all  $4^{N_q}$ Pauli strings we denote $\cal{P}$. These Pauli strings form an orthonormal basis for Hermitian operators on the system such that
\begin{equation}
\Braket{ \hat{\sigma}_s | \hat{\sigma}_{s'}} = \frac{1}{\mathcal{N}}\Trace\left(\hat{\sigma}_s \hat{\sigma}_{s'} \right) =  \delta_{s,s'}.
\label{eq:orthocondition}
\end{equation}

\section{Calculating the Lieb-Robinson correlation function \label{sec:PauliWalkMethod}}

\subsection{Heisenberg time dependence}
In the Heisenberg picture, the time dependence of any operator can be written 
\begin{equation}
\hat{Q}(t) = \sum_{n=0}^{\infty} \frac{1}{n!} \left(\frac{it}{\hbar} \right)^n \left[\left(\hat{H}\right)^{n},\;\hat{Q}\right]
\label{eq:heisenberg_operator}
\end{equation}
where $\hat{Q}=\hat{Q}(0)$, and we use the usual notation for $n$ 
\begin{myhighlight}
iterated commutators
\begin{equation}
\left[ \left(A \right)^n, B \right] \equiv 
[ \underbrace{A,\dots [A, [ A}_{\text{n times}}, B ] \dots] ]
\label{eq:nested_commutation}
\end{equation}
\end{myhighlight}


\noindent Using Eq.~(\ref{eq:tauDef}), the time dependence of $\hat{\sigma}^z_1$ can then be written
\begin{equation}
\sigma^z_{1}(t) = \sum_{n=0}^{\infty} \frac{1}{n!} \pi^n i^n \left(\frac{t}{\tau} \right)^n \left[\left(\hat{H}'\right)^{n},\sigma_1^{z}\right].
\label{eq:sigmaz1Oft}
\end{equation}

\noindent The Lieb Robinson correlation function is  defined by
\begin{align}
C_k(t) &\equiv \| \left[\hat{\sigma}_1^z, \hat{\sigma}_k^z(t) \right]  \|
=\| \left[\hat{\sigma}_k^z,\hat{\sigma}_1^z (t) \right]  \| \nonumber\\
&= \left\| \left[\hat{\sigma}_k^z,\; 
e^{i\hat{H}t/\hbar} \hat{\sigma}_1^z  e^{-i\hat{H}t/\hbar}  \right]\right\|
\label{eq:LRexponentialDef}
\end{align}
and we take this as the most direct method of computing it.
Making explicit use of the norm defined in \refeq{NormDef} we have
\begin{align}
C_k(t)&=\sqrt{\Trace \left(\frac{\left|\left[\hat{\sigma}_k^z,\hat{\sigma}_1^z (t) \right] \right|^2}{\cal{N}} \right)}.
\label{eq:LRcorrelationFunctionDef}
\end{align}

From the definition, we see $C_k(0) = 0$ because $\hat{\sigma}_1^z$ commutes with itself and with the Pauli operators on all other sites. The time dependence of $\hat{\sigma}^z_1(t)$ can be envisioned as  the operator spreading out from the first qubit progressively down the array, incorporating components of Pauli operators on other qubits. As the expanding operator encounters site $k$, the quantum correlation between the qubit $1$ and qubit $k$, as quantified by $C_k(t)$, increases. It is this spread and growth that we want to understand in detail.

We focus first on the operator formed by the commutator
$\left[\hat{\sigma}_k^z,\hat{\sigma}_1^z (t) \right]$ and using Eq.~(\ref{eq:sigmaz1Oft}) obtain

\begin{equation}
\left[\hat{\sigma}_k^z,\hat{\sigma}_1^z (t) \right] = \sum_{n=0}^{\infty} \frac{1}{n!} \pi^n i^n \left(\frac{t}{\tau} \right)^n \left[\hat{\sigma}_k^z,\left[\left( \hat{H}'\right)^{n},\sigma_1^{z}\right]\right].
\label{eq:CommutatorExpansion1}
\end{equation}

\noindent Because Pauli strings span the operator space, the commutator can be expanded as a weighted sum of Pauli strings as

\begin{equation}
\left[\left( \hat{H}' \right)^n, \hat{\sigma}^z_1 \right] =
\sum_{\hat{\sigma}_s \in \cal{P}} C_{n,\hat{\sigma}_s} \hat{\sigma}_s.
\label{eq:Cnsigmadefine}
\end{equation}

\noindent To solve for the coefficients $C_{n,\hat{\sigma}s} $, we  multiply both sides of (\ref{eq:Cnsigmadefine}) by $\hat{\sigma}_{s'}$ 

\begin{align}
\left[\left( \hat{H}'\right)^n, \hat{\sigma}^z_1 \right] \hat{\sigma}_{s'} = \left(
\sum_{\hat{\sigma}_s \in \cal{P}} C_{n,\hat{\sigma}_s} \hat{\sigma}_s \right) \hat{\sigma}_{s'},
\end{align}
and take the trace
\begin{align}
\Trace\left(\left[\left( \hat{H}'\right)^n, \hat{\sigma}^z_1 \right] \hat{\sigma}_{s'}\right)= \sum_{\hat{\sigma}_s \in \cal{P}}C_{n,\hat{\sigma}_s} \Trace\left(
\hat{\sigma}_s\hat{\sigma}_{s'} \right).
\label{TracenormResult1}
\end{align}

\noindent Now using Eq.~(\ref{eq:orthocondition}), we obtain

\begin{align}
C_{n,\hat{\sigma}_s} = \frac {\Trace \left(\left[\left( \hat{H}'\right)^{n},\sigma_1^{z}\right] \hat{\sigma}_s \right)}{\mathcal{N}} 
\end{align}
or
\begin{align}
C_{n,\hat{\sigma}_s} =\Braket{ \left[\left( \hat{H}'\right)^{n},\sigma_1^{z}\right] | \hat{\sigma}_s    }.
\label{eq:Cexpression}
\end{align}

\noindent By substituting  Eq.~(\ref{eq:Cnsigmadefine}) into  Eq.~(\ref{eq:CommutatorExpansion1}), we have

\begin{equation}
\left[\hat{\sigma}_k^z,\hat{\sigma}_1^z (t) \right] = \sum_{\hat{\sigma}_s \in \cal{P}} \sum_{n=0}^{\infty} \frac{1}{n!} \pi^n \left(i\right)^n 
\left(\frac{t}{\tau} \right)^n C_{n,\hat{\sigma}_s}\left[\hat{\sigma}_k^z,\hat{\sigma}_s \right].
\label{eq:doubleSummation}
\end{equation}

\noindent  We define

\begin{equation}
\hat{\sigma}_{k,s} \equiv \frac{1}{2} \left[\hat{\sigma}^z_k, \hat{\sigma}_s \right].
\label{eq:sigmaks}
\end{equation}


By virtue of \refeq{orthocondition}, one can show that the operators $\hat{\sigma}^\dagger_{k,s}$ also have an orthogonality relationship,
\begin{equation}
    \frac{1}{\mathcal{N}} \Tr \left( \hat{\sigma}^\dagger_{k,s} \hat{\sigma}_{k,s'}   \right) =
    \delta_{s,s'} \frac{1}{\mathcal{N}} \Tr \left( \hat{\sigma}^\dagger_{k,s} \hat{\sigma}_{k,s} \right).
    \label{eq:SigmaksOrthogonality}
\end{equation}
From \refeq{doubleSummation} can write
\begin{equation}
\left[\hat{\sigma}_k^z,\hat{\sigma}_1^z (t) \right] =\sum_{\hat{\sigma}_s \in \cal{P}} \sum_{n=0}^{\infty} \frac{2}{n!} \pi^n \left(i\right)^n \left(\frac{t}{\tau} \right)^n C_{n,\hat{\sigma}_s} \hat{\sigma}_{k,s}.
\label{eq:commutatorsigmak}
\end{equation}

It is necessary to calculate the trace of the absolute value square of this commutator so that we can evaluate the Lieb Robinson correlation function using Eq.~(\ref{eq:LRcorrelationFunctionDef}). We proceed by defining  two quantities that will be temporarily useful,  

\begin{equation}
R_n\left(t\right) \equiv \frac{2}{n!} \pi^n \left(i\right)^n \left(\frac{t}{\tau} \right)^n,
\label{eq:Rn1}
\end{equation}

\noindent and

\begin{equation}
B_{\hat{\sigma}_s}(t) \equiv \sum_{n=0}^{\infty} R_n\left(t\right)C_{n,\hat{\sigma}_s}.
\label{eq:Bsigmas1}
\end{equation}

\noindent Note that both the  quantities $R_n(t)$ and $B_{\hat{\sigma}_s}(t)$ are numbers rather than operators. Using Eqs.~(\ref{eq:Rn1}) and (\ref{eq:Bsigmas1}), we can rewrite  Eq.~(\ref{eq:commutatorsigmak}) in the compact form

\begin{equation}
\left[\hat{\sigma}_k^z,\hat{\sigma}_1^z (t) \right] =  \sum_{\hat{\sigma}_{s'} \in \cal{P}} B_{\hat{\sigma}_s'}(t) \hat{\sigma}_{k,s'}.
\label{eq:commutatorSum}
\end{equation}

\noindent The absolute value squared of the commutator is then

\begin{align}
\left|\left[\hat{\sigma}_k^z,\hat{\sigma}_1^z (t) \right] \right|^2 &= \left( \sum_{\hat{\sigma}_s \in \cal{P}}B_{\hat{\sigma}_s}(t) \hat{\sigma}_{k,s}\right)^\dagger \nonumber\\
&\times
\left( \sum_{\hat{\sigma}_{s'} \in \cal{P}} B_{\hat{\sigma}_{s'}}(t) \hat{\sigma}_{k,s'}\right) \\
&= \sum_{\hat{\sigma}_{s},\hat{\sigma}_{s'}\in \cal{P}} B^*_{\hat{\sigma}_{s}} \left(t\right) \hat{\sigma}_{k,s}^\dagger B_{\hat{\sigma}_{s'}} \left( t\right) \hat{\sigma}_{k,s'}.
\label{eq:squaredCommutator}
\end{align}
so
\begin{align}
\frac{1}{\mathcal{N}}\Trace\left(\left|\left[\hat{\sigma}_k^z,\hat{\sigma}_1^z (t) \right] \right|^2\right) &= 
\sum_{\hat{\sigma}_s, \hat{\sigma}_{s'} \in \mathcal{P}} B^*_{\hat{\sigma}_{s}}(t)  B_{\hat{\sigma}_{s'}}(t)\nonumber\\
&\times
\frac{ \Trace \left( \hat{\sigma}_{k,s}^\dagger \hat{\sigma}_{k,s'} \right)}{\mathcal{N}} \\
&=  \sum_{\hat{\sigma}_s \in \mathcal{P}}  \left| B_{\hat{\sigma}_{s}}(t) \right|^2  D_{k,\sigma_s}
\label{Trace1LR}
\end{align}
where we have used \refeq{SigmaksOrthogonality} and define 
\begin{align}
D_{k,\sigma_s} \equiv \Braket{\hat{\sigma}_{k,s}| \hat{\sigma}_{k,s} }&= \frac{\Trace\left(\hat{\sigma}_{k,s}^\dagger \hat{\sigma}_{k,s}\right)}{\mathcal{N}} \nonumber\\
&= \left\| \frac{1}{2} \left[ \hat{\sigma}_k^z, \hat{\sigma}_s \right] \right\|^2 .
\label{eq:DefineDks}
\end{align}


\noindent Re-inserting the expressions for $B_{\hat{\sigma}_{s}}(t)$ and $R_n(t)$ from Eqs.~(\ref{eq:Rn1}) and (\ref{eq:Bsigmas1}) yields
\begin{multline}
\frac{1}{\mathcal{N}} \Trace\left(\left|\left[\hat{\sigma}_k^z,\hat{\sigma}_1^z (t) \right] \right|^2 \right) \\
=\sum_{\sigma_s \in \cal{P}} \left| \left(\sum_{n=0}^{\infty} \frac{2}{n!} \pi^n \left(i\right)^n \left(\frac{t}{\tau} \right)^nC_{n,\hat{\sigma}_s}\right)  \right|^2 D_{k,\sigma_s}, 
\label{eq:keyequationTrace}
\end{multline}
and so from the definition for the correlation function in Eq.~(\ref{eq:LRcorrelationFunctionDef}), we have
\begin{equation}
    \displaystyle{C_k(t) = \sqrt{\displaystyle{\sum_{\sigma_s \in \cal{P}}} \left| \left(\sum_{n=0}^{\infty} 
    \frac{2}{n!} 
     \left(i \pi \frac{t}{\tau} \right)^n
    C_{n,\hat{\sigma}_s}\right)  \right|^2 D_{k,\sigma_s}}}
    \label{eq:LRcorrCandD}
\end{equation}

To calculate this correlation function we need to determine the quantity $C_{n,\hat{\sigma}_s}$  from Eq.~(\ref{eq:Cexpression}) and $D_{k,\sigma_s}$ from Eq.~(\ref{eq:DefineDks}).

\subsection{Calculating  $C_{n,\hat{\sigma}_s}$ using operator Pauli walks}

In this section, we develop an algorithm for calculating $C_{n,\hat{\sigma}_s}$ using algebraic methods, graph theory, and finally matrix methods. Specifically, we aim to determine the subset of the $4^{N_q} $ Pauli strings that actually contribute to the sum in (\ref{eq:LRcorrCandD}), and to calculate the inner products in  \refeq{Cexpression}.

We now write the Hamiltonian in the more compact notation
\begin{equation}
    H' = -\sum_{j=1}^{N_q} X_j  -
    J' \sum_{j=1}^{N_q-1} Z_j Z_{j+1}.
\end{equation}

We need to calculate the projection of Pauli string $\hat{\sigma}_s$ on the iterated commutator
\begin{align}
C_{n,\hat{\sigma}_s} &= \Braket{ \left[\left( H'\right)^{n},Z_1\right] | \hat{\sigma}_s}
\label{eq:CexpressionInnerProduct}\\
&= \frac{1}{\mathcal{N}} \Tr \left( \left[\left( H'\right)^{n},Z_1\right]   \hat{\sigma}_s \right)
\label{eq:CexpressionInnerProductTrace}
\end{align}

\subsection{Example: Four qubit case}

We first consider a chain of $N_q=4$ qubits and then will generalize later. 
Our notation for Pauli strings will suppress identity operators so, for example, $X_1 Y_2$ will be understood to mean $X_1 Y_2 I_3 I_4$.  
We focus first on the iterated commutator 
\begin{equation}
    \left[\left(H'\right)^{n},Z_1\right].
    \label{eq:IteratedHprimeComm}
\end{equation}
We begin by calculating 
$\left[H',Z_1\right]$ and generate the other commutators that arise as a result of the repeated iteration of commutation with $H'$ in \refeq{CexpressionInnerProductTrace}.
The Pauli operator commutation relations yield 

\begin{align}
 \left[H',Z_1\right] &= 2i \underline{Y_1}   \label{eq:HprimeFirstTerm} \\
 \left[H',\underline{Y_1}\right] &= -2i Z_1 + 2iJ’\underline{\underline{X_1 Z_2}} \label{eq:HprimeSecondTerm} \\
 \left[H',\underline{\underline{X_1 Z_2}}\right] &= 2i \underline{X_1 Y_2} - 2i J'Y_1 \label{eq:HprimeThirdTerm} \\
 \left[H',\underline{X_1 Y_2}\right]  &= -2i X_1 Z_2 + 2i J'\underline{\underline{X_1 X_2 Z_3}} \label{eq:HprimeFourthTerm} \\
\left[H',\underline{\underline{X_1 X_2 Z_3}}\right] &= 2i \underline{X_1 X_2 Y_3} - 2i J'X_1 Y_2 \label{eq:HprimeFifthTerm} \\
\left[H',\underline{X_1 X_2 Y_3}\right]
    &= -2i X_1 X_2 Z_3 + \nonumber \\ 
    &\quad 2i J'\underline{\underline{X_1 X_2 X_3 Z_4}} \label{eq:HprimeSixthTerm} \\
  \left[H',\underline{\underline{X_1 X_2 X_3 Z_4}}\right] &=2i \underline{X_1 X_2 X_3 Y_4} - 2i J'X_1 X_2 Y_3 \label{eq:HprimeSeventhTerm} \\
      \left[H',\underline{X_1 X_2 X_3 Y_4}\right]  &=  -2i X_1 X_2 X_3 Z_4.
      \label{eq:HprimeLastTerm}
\end{align}
In each of these equations, the term underlined on the right gives us the next commutator to evaluate. For each equation, the commutator between $H'$ and the non-underlined term can be evaluated using the previous results.  The sequence terminates  in his case because there is no fifth qubit.

One result of this process is that the sequence allows us to order the relevant Pauli strings   as
\begin{align}
\sigma_1 &=\quad Z_1  \label{eq:Paulilabel1} \\
\sigma_2 &=\quad Y_1 \\
\sigma_3 &=\quad X_1 Z_2 \\
\sigma_4 &=\quad X_1 Y_2\\
\sigma_5 &=\quad X_1 X_2 Z_3 \\
\sigma_6 &=\quad X_1 X_2 Y_3 \\
\sigma_7 &=\quad X_1 X_2 X_3 Z_4 \\
\sigma_8 &=\quad X_1 X_2 X_3 Y_4 \label{eq:Paulilabel2}.
\end{align}
The Pauli strings $\sigma_m$  enumerated in  Eqs.~(\ref{eq:Paulilabel1})--(\ref{eq:Paulilabel2})   form a closed set in the sense that the iterated commutator $\left[\left(\hat{H}'\right)^{n},Z_1\right] $ will be a linear combination of these Pauli strings and no others for all values of $n$. 
All other Pauli strings  $\hat{\sigma}_s$ will yield an inner product $C_{n,\hat{\sigma}_s}=0$. Rather than considering all $4^4=256$ Pauli strings,  we will only need to consider $C_{n,\hat{\sigma}_m}$ for $m=[1,8]$.


Next we can use Eqs.~(\ref{eq:HprimeFirstTerm})-(\ref{eq:HprimeLastTerm}) to calculate the iterated commutators in Eq.~(\ref{eq:IteratedHprimeComm}) and the inner products in (\ref{eq:CexpressionInnerProduct}). We  evaluate the first few using operator algebra. For the $n=1$ iterated commutator we have

\begin{align}
    \left[(H')^1,Z_1\right] &= 2i Y_1,
\end{align}
therefore from \refeq{CexpressionInnerProduct}
\begin{equation}
    C_{1,Y_1}=2i
\end{equation}
The next level of iteration can be evaluated using  \refeq{HprimeSecondTerm} to obtain
\begin{align}
    \left[(H')^2,Z_1\right] &= 2i \left[H', Y_1  \right]\\
                            &= (2i) (-2i) Z_1 + (2i)(2i J') X_1 Z_2 
\end{align}
so evaluating the inner product in \refeq{CexpressionInnerProduct} yields
\begin{align}
    C_{2,Z_1}&= (2i)(-2i) = 4 \\
    C_{2,X_1 Z_2} &= (2i)(2iJ')= -4 J'.
\end{align}
Similarly
\begin{align}
    \left[(H')^3,Z_1\right] &= \left[\; (2i)(2i)(-2i) Y_1 \right. \nonumber\\
                             & \qquad +(2i)(2iJ')(2i) X_1 Y_2  \nonumber\\
                            & \qquad \left . + (2i)(2i J')(-2iJ') Y_1 \; \right]
\end{align}
so 
\begin{align}
    C_{3,Y_1}&= \left( (2i)(2i)(-2i) + (2i)(2i J')(-2iJ') \right)\\ 
    &=8 i + 8i(J')^2   \label{eq:C3Y1}\\
    C_{3,X_1 Y_2} &= (2i)(2iJ')(2i)\\
    &= -8 iJ' .
    \label{eq:C3X1Y2}
\end{align} 
To evaluate the sum in \refeq{LRcorrCandD}, we need to evaluate  $C_{n,\hat{\sigma}_m}$ for arbitrarily large values of $n$.

\begin{figure}[bt]
\centering
\includegraphics[width=\columnwidth]{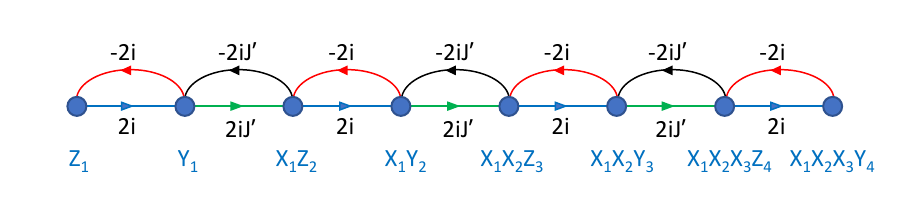 }
\caption{Operator node graph for a  4 qubit line.  
}
\label{fig:EightNodeSchematic}
\end{figure} 

We  now construct a finite directed graph which represents this algebraic process. Figure \ref{fig:EightNodeSchematic} shows an array of operator {\em nodes} (in blue) corresponding to the 8 Pauli strings listed in (\ref{eq:Paulilabel1})--(\ref{eq:Paulilabel2}). 
The connectivity of the nodes is determined by Eqs.~(\ref{eq:HprimeFirstTerm})-(\ref{eq:HprimeLastTerm}).
Two  nodes are connected by an {\em edge} if the two corresponding operators appear on the left and right side of one of these equations. The {\em direction} of the edge (indicated in the figure by an arrow) is from the operator that appears on the left side of the equation to the operator that appears on the right side. 
The {\em edge weight}  is given by the coefficient of the operator on the right side of the equation.
Edges are color-coded in Figure \ref{fig:EightNodeSchematic} according to their weight and direction.

For example, consider the node in Figure \ref{fig:EightNodeSchematic} corresponding to the operator $X_1 Z_2$. There are four edges connecting this node to other nodes:
\begin{itemize}
    \item The right-directed edge going from node $ Y_1$ with weight $2iJ'$ given by \refeq{HprimeSecondTerm}
    
    \item The right-directed edge going to node $X_1 Y_2$ with weight $2i$ given by \refeq{HprimeThirdTerm}

    \item The left-directed edge going to node $ Y_1$ with weight $-2iJ'$ given by \refeq{HprimeThirdTerm}

     \item The left-directed edge going from node $X_1 Y_2$ with weight $-2i$ given by \refeq{HprimeFourthTerm}.

\end{itemize}
 
We define an  {\em operator Pauli walk} on this graph as an ordered list of nodes (Pauli strings) such that each consecutive pair of nodes is connected by a directed edge.  Each node may be visited more than once and each edge traversed more than once. The {\em length} of  such a Pauli walk is the number of edge traversals, or equivalently, one less than the number of (not necessarily unique) nodes visited along the walk. The {\em  weight product} of a Pauli walk is the product of the edge weights along the walk \cite{Osborn2010}. 

The sequence of iterated commutators generated by Eqs.~(\ref{eq:HprimeFirstTerm})-(\ref{eq:HprimeLastTerm}) correspond precisely to a Pauli walk among the nodes associated with the Pauli strings $\sigma_m$ in Eqs.~(\ref{eq:Paulilabel1})--(\ref{eq:Paulilabel2}).
The quantity $ C_{n, \sigma_m}$ is equal to the sum of the  weight products for all Pauli walks of length $n$ which start at the first node ($\hat{\sigma}_1=Z_1$) and ends at the node labeled $\hat{\sigma}_m$.

Some examples:
\begin{itemize}

    \item There is 1  Pauli walk of length 3 connecting node $Z_1$ to node $X_1 Y_2$:

\hspace{0.5in} $\{Z_1, Y_1, X_1 Z_2, X_1 Y_2\}$. 

From Figure~\ref{fig:EightNodeSchematic} we see that the weight product for this walk is  
\begin{equation}
    C_{3,X_1 Y_2} =  (2i)(2iJ')(2i)=-8 i J'
\end{equation} 
in agreement with \refeq{C3X1Y2}.

\item There are 2 Pauli walks of length 3 connecting node $Z_1$ to node $Y_1$:

\hspace{0.25in} $\{Z_1, Y_1, Z_1, Y_1\}$ and 
$\{Z_1, Y_1, X_1 Z_2,  Y_1\}$.

From the figure we see that the sum of the corresponding weight products is
\begin{align}
C_{3,X_1 Y_2} &= \left[ (2i)(-2i)(2i) \right. \\
& \quad + \left. (2i)(2i J')(-2iJ') \right] \\
&= 8i + 8iJ'
\end{align}
in agreement with \refeq{C3Y1}.

\item There are 3 Pauli walks of length 4 connecting node $Z_1$ to node $X_1 Z_2$:
\begin{align}
&\{Z_1, Y_1, X_1 Z_2, X_1 Y_2, X_1 Z_2 \}, \\
&\{Z_1, Y_1, Z_1, Y_1, X_1 Z_2 \},\quad \text{and} \\
&\{Z_1, Y_1, X_1 Z_2, Y_1, X_1 Z_2 \}.
\end{align}

From the figure we see that the sum of the corresponding  weight products is 
\begin{align}
         C_{3,X_1 Z_2} &=  \left[ (2i) (2iJ')  (2i) (-2i) \right. \nonumber\\
         &\quad + (2i) (-2i) (2i) (2iJ')  \nonumber\\
         &\quad + (2i) (2iJ') (-2iJ') (2iJ') \left. \right] \nonumber\\ 
         &= -16 \left(J'\right)^3- 32 J'.
\end{align}

\end{itemize}
In general, $C_{n,\hat{\sigma}_m}$ is a polynomial in $J'$. As the walk length $n$ becomes large, the number of walks becomes very large, even for a modest $N_q$, because each walk can go back and forth between nodes many times. 

The process of computing the sum of the weight products for all Pauli walks of a given  length can be automated by defining the adjacency matrix $\bm{A}$ whose $(\ell, p)$ element is the edge weight from the $\ell^{\text{th}}$ node to the $p^{\text{th}}$ node. In the 4-qubit example  illustrated in Figure~\ref{fig:EightNodeSchematic}  the $8\times 8$ adjacency  matrix is

\begin{equation}
\bm{A} = 2i\begin{pmatrix}
0   & 1    & 0    & 0     & 0    & 0   & 0   & 0  \\
-1  & 0    & J'   & 0     & 0    & 0   & 0   & 0  \\
 0  &-J'   & 0    & 1     & 0    & 0   & 0   & 0  \\
0   & 0    & -1   & 0     & J'   & 0   & 0   & 0  \\
0   & 0    & 0    &-J'    & 0    & 1   & 0   & 0  \\
0   & 0    & 0    & 0     & -1   & 0   & J'  & 0  \\
0   & 0    & 0    & 0     & 0    &-J'  & 0   & 1 \\
0   & 0    & 0    & 0     & 0    & 0   & -1   & 0  
\end{pmatrix}.
\end{equation}
It is convenient to also define $\bm{A'}\equiv \bm{A}/(2i)$.
The sum of the weight products of all walks of length $n$ that begin at node $j$ and end at node $k$  is the  $(j,k)^{\text{th}}$ element of the  $n^{\text{th}}$ power of $\bm{A}$. In our case, this means we can  evaluate the sum of all weight products for Pauli walks  starting at the first node $\hat{\sigma}_1=Z_1$ and ending at node $\hat{\sigma}_m$ by evaluating
\begin{equation}
    C_{n,\sigma_m} =\left(\bm{A}^n\right)_{1,m}.
    \label{eq:GnmFromA}
\end{equation}

Using  (\ref{eq:GnmFromA}), we can now rewrite \refeq{LRcorrCandD} with the sum over the set of Pauli strings $\sigma_s$ now expressed as a sum over the ordered set $\{\sigma_m\}$  as

\begin{equation}
    C_k(t) = \sqrt{ \sum_{m=1}^{8} 
    \left| \sum_{n=0}^{\infty} \frac{2}{n!} \left(-2\pi \frac{t}{\tau} \right)^n \left[ (\bm{A}^')^n \right]_{1,m}  \right|^2 D_{k,m}}
    \label{eq:LRcorrAandDmNq4}
\end{equation}
where 
\begin{equation}
D_{k,m} = \Braket{\hat{\sigma}_{k,m}| \hat{\sigma}_{k,m} } = \left\| \frac{1}{2} \left[ Z_k, \hat{\sigma}_m \right] \right\|^2 .
\label{eq:DefineDkm}
\end{equation}

\subsection{Calculating $D_{k,m}$}

We have seen that for the 4-qubit line, only the $8$ Pauli strings  enumerated in  Eqs.~(\ref{eq:Paulilabel1})--(\ref{eq:Paulilabel2})  contribute to the correlation function in \refeq{LRcorrCandD}. Examining the structure of this set, we see that the Pauli strings in $\{ \sigma_m\}$ appear in pairs associated with each qubit $k$. The first element of each pair ends  with $Z_k$, and the second ends with with $Y_k$, after which all the succeeding $\sigma_m$'s have the operator $X_k$.  The quantity $D_{k,m}$ is zero for any Pauli string which includes $Z_k$ or $I_k$ because both commute with $Z_k$. Only if $\sigma_m$ includes $Y_k$ or $X_k$ is $D_{k,m}\ne 0$, and in both of those cases it has a value of  $1$. The result is that $D_{k,m}$ is 0 for $m < 2k $ and 1 otherwise. Therefore the effect of $D_{k,m}$ is  to limit the sum over $m$ in \refeq{LRcorrAandDmNq4} to start at $m=2k$,  so we can write


\begin{equation}
    C_k(t) = \sqrt{ \sum_{m=2k}^{8} 
    \left| \sum_{n=0}^{\infty} \frac{2}{n!} \left(-2\pi \frac{t}{\tau} \right)^n \left[ (\bm{A}^')^n \right]_{1,m}  \right|^2 }.
    \label{eq:LRcorrANq4}
\end{equation}


\subsection{Generalizing the method of operator Pauli walks}

Generalizing from the case of $N_q=4$ to any number of qubits is now straightforward. 
In place of the 8 equations relations in Eqs.~(\ref{eq:HprimeFirstTerm})-(\ref{eq:HprimeLastTerm}), we have $2N_q$ commutation relations. For each qubit with index $j=1$ to $N_q$, there are two relevant equations: one is the commutator between the Hamiltonian and  a Pauli string that ends in $Z_j$, and the second is a commutator between the Hamiltonian and  a Pauli string that ends in $Y_j$. Specifically, we have
\begin{multline}
   \left[{H}',\left(\prod_{k = 1}^{j-1} X_k \right)Z_j\right] =  2i \left(\prod_{k = 1}^{j-1} X_k \right) Y_j \\
   - 2iJ'\left(\prod_{k = 1}^{j-2} X_{k} \right) Y_{j-1}, 
   \quad 1 \le j \le N_q \\
\end{multline}
and
\begin{multline}
\left[{H}',\left(\prod_{k = 1}^{j-1} X_k \right)Y_j\right] =\hfill\\
\qquad\qquad\begin{cases}
    { -2i \left(\prod\limits_{k = 1}^{j-1} X_k \right) Z_j  }\\
    \quad +2iJ'\left(\prod\limits_{k = 1}^{j} X_{k} \right) Z_{j+1}, 
     \;1 \le j < N_q   \\
    {-2i \left(\prod\limits_{k = 1}^{j-1} X_k \right) Z_j}, \hfill j = N_q.
     \end{cases} \hfill
\label{eq:HcommutatorGeneral2}
\end{multline}
These connect the  Pauli strings that contribute to \refeq{LRcorrCandD}. These $2N_q$ operators $\sigma_m$  also occur in pairs associated with qubit  index $j\in [1,N_q]$.
\begin{align}
    \sigma_1 &= Z_1  \nonumber \\
    \sigma_2 &= Y_1  \nonumber \\
    \sigma_{2j-1} &= \left(\prod_{k=1}^{(j-2)/2} X_k\right) Y_{(j/2)}, \quad 
                         &j \text{ even, } j \ge 2  \nonumber\\
    \sigma_{2j} &=   \left(\prod_{i=1}^{(j-1)/2} X_i \right)Z_{(j+1)/2}, \quad
                          &j \text{ odd, } j \ge 2 
\label{eq:SigmaMLongChain}           
\end{align}

The method of enumerating Pauli walks on the nodes labeled by  $\sigma_m$ as described in the previous section is naturally extended to any length chain. 
The adjacency matrix $\bm{A} $, determined by the coefficients in Eqs.~(\ref{eq:HcommutatorGeneral2}) and (\ref{eq:HcommutatorGeneral2}), can  be written in the general case as
\begin{equation}
\bm{A} = 2i\begin{pmatrix}
0      & 1     & 0      & 0      & 0       & 0      & \cdots   & 0  \\
-1     & 0     & J'     & 0      & 0       & 0      & \cdots   & 0  \\
 0     &-J'    & 0      & 1      & 0       & 0      & \cdots   & 0  \\
0      & 0     & -1     & 0      &J'       & 0      & \cdots   & 0  \\
0      & 0     & 0      &-J'     & 0       & \ddots & \vdots   & \vdots\\
0      & 0     & 0      & 0      & \ddots  & \ddots & J'       & 0  \\
\vdots &\vdots &\vdots  & \vdots & \cdots  &-J'     & 0        & 1 \\
0      & 0     & 0      & 0      &\cdots   & 0      & -1       & 0  
\end{pmatrix}=2i\bm{A}'
\label{eq:AdjacencyArbNq}
\end{equation}
and we  again define $\bm{A'}\equiv \bm{A}/(2i)$.

The value of the Lieb-Robinson correlation function can be calculated by calculating powers of $\bm{A}^'$ and evaluating 

\begin{equation}
    C_k(t) = \sqrt{ \sum_{m=2k}^{2N_q} 
    \left| \sum_{n=0}^{\infty} \frac{2}{n!} \left(-2\pi \frac{t}{\tau} \right)^n \left[ (\bm{A}^')^n \right]_{1,m}  \right|^2 }.
    \label{eq:LRcorrArbSumForm}.
\end{equation}
\begin{myhighlight}  
This can be rewritten in terms of the matrix exponential of the adjacency matrix:
\begin{equation}
     C_k(t) = \sqrt{\sum_{m = 2k}^{2N_q} \left| 2 \left[ \exponential{\left(-2 \pi \frac{t}{\tau}\bm{A}^'\right)}\right]_{1,m} \right|^2}.
    \label{eq:LRcorrArb}
\end{equation}       

\end{myhighlight}
Importantly, whereas the size of the Hamiltonian matrix for a system of $N_q$ is 
$2^{N_q} \times  2^{N_q} $,  the size of the connectivity matrix $\bm{A}$ is only $2N_q \times 2N_q$. This makes the method tractable for much larger systems. To evaluate $C_q(t)$ with this method only requires evaluating \refeq{LRcorrArb}.

\begin{myhighlight}  
\subsection{Choosing a norm \label{sec:Norms}}
We can now revisit the question of which norm to use in computing the Lieb-Robinson correlation function. 

The operator norm and Frobenius norm of an operator $Q$ can be expressed as
\begin{eqnarray}
    \| Q \|_{\text{Op}} &=& \sqrt{ \lambda_{\text{max}} 
        \left(  \op{Q}\hc{Q}   \right) } \\
    \| Q \|_{\text{Frob}} &=& \sqrt{ \frac{1}{\cal{N}}\Tr \left( \op{Q}\hc{Q}  \right) }.
\end{eqnarray}
The operator norm is the square root of the maximum eigenvalue of $QQ^\dagger$ and the Frobenius norm is the square root of the average of the eigenvalues of $QQ^\dagger$. 
In general these are of course not equal to each other. 

For the QTFIM,  we have shown above that the operator $Q= \left[\hat{\sigma}_k^z,\hat{\sigma}_1^z (t) \right]$ can be written as a linear combination of just the $2N_q$ Pauli strings in \refeq{SigmaMLongChain}. One can show that therefore in this case, for all times $QQ^\dagger$ is a (time-dependent) multiple of the identity operator. For the identity operator $I$, the eigenvalues are all $1$, so the largest eigenvalue and the average eigenvalue are identical.
The two norms then yield the same result. 
The detailed proof is  included as Supplemental Material \cite{SupplementalMaterial}.  As a consequence of this identity, the Lieb-Robinson correlation function defined by \refeq{LRcorrDef} can be calculated with either definition of the norm; we have chosen to use the Frobenius norm.
\end{myhighlight}  

\subsection{Comparison with the direct time-exponential method}
\begin{figure}[tb]
\centering
    \includegraphics[width=\columnwidth]{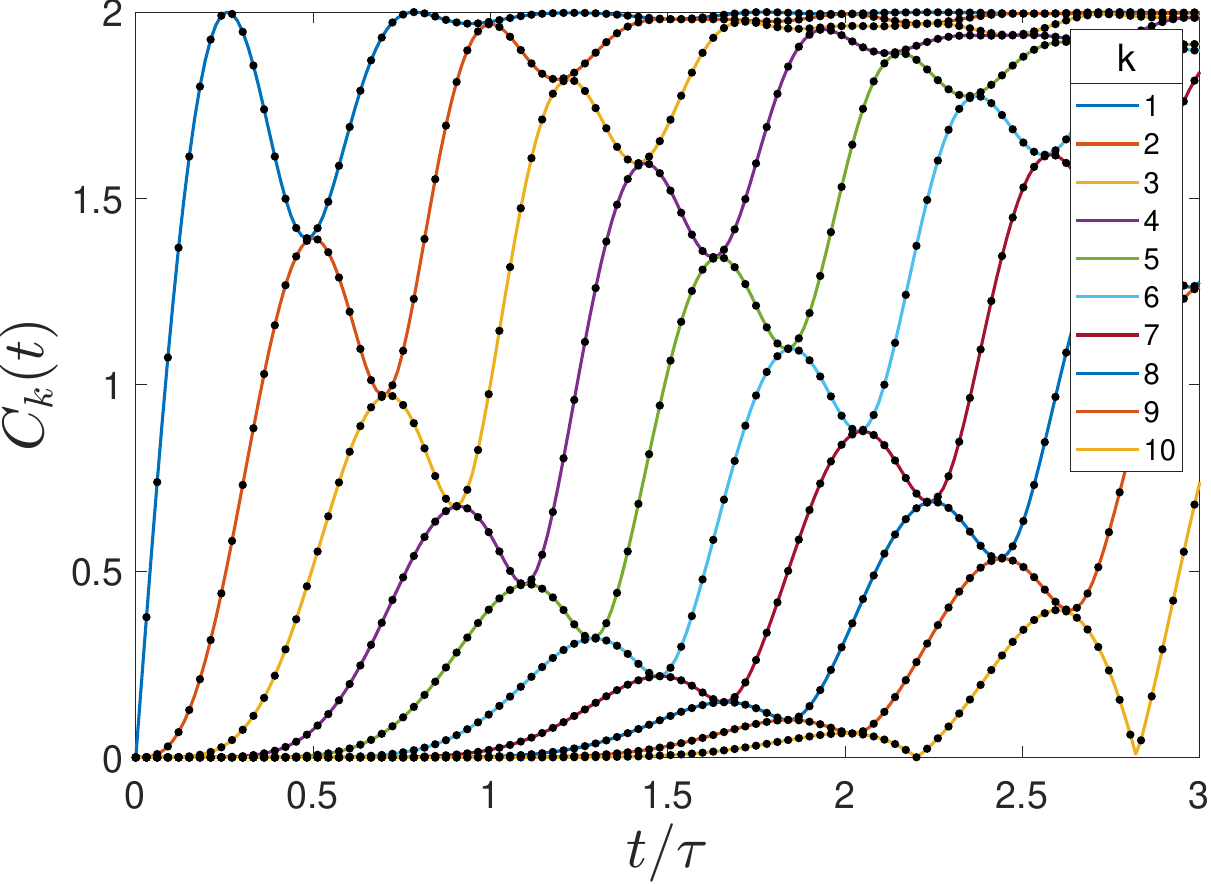}
    \caption{The Lieb-Robinson correlation function for a ten-qubit chain with $J'=0.5$. The solid curve shows the results for $C_k(t)$ calculated using the exponential time evolution operator in \refeq{LRexponentialDef}. The points show the results calculated from the method of Pauli walks using \refeq{LRcorrArb}. The Pauli walk method in this case is  more than 100 times faster and can be extended to much larger chains.
    }
    \label{fig:CompareDirectWalkShort}
\end{figure}

Figure \ref{fig:CompareDirectWalkShort} shows the Lieb-Robinson correlation function $C_k(t)$ for a chain of $N_q=10$ qubits. The timescale is in units of $\tau$ as defined by \refeq{tauDef}. The solid curves show the results of a direct calculation using the operator exponential time dependence in \refeq{LRexponentialDef}.The points show the result of the operator Pauli walk method using \refeq{LRcorrArb}.
The agreement between the two methods is essentially exact. The results are shown for  qubit coupling  $J'=1/2$. Because large-matrix exponentials are costly to calculate,  the Pauli walk method here provides a speed-up of more than a factor of 100  for the calculation. To assure accuracy, we use an extended-precison arithmetic package \cite{AdvanpixMultiprecision}.

Several general features of $C_k(t)$ are visible even in this relatively short line. The value of the correlation function is zero at $t=0$ for all $k$ because operators on the first qubit and the $k^{\text{th}}$ qubit initially commute---they cannot know about each other yet.  As the influence of the first qubit propagates down the chain,  the correlation function for each site starts  to rise.  Quantum oscillations on the characteristic time scale of $\tau$ are due to the internal dynamics of each qubit. The maximum value of $C_k(t)$ is $C_{\text{sat}}=2$. This has its origin in the Pauli commutation relations of \refeq{PauliCommRelation}---the maximum amount of "non-commutation" is 2.  The correlation of qubit 1 with itself, $C_1(t)$ is  initially zero because 
$[\hat{\sigma}_1^z(0),  \hat{\sigma}_1^z]=0$, and rises because $\hat{\sigma}_1^z(t)$ starts to mix in components of $\hat{\sigma}_1^x$ and $\hat{\sigma}_1^y$.

The nesting of $C_k(t)$ for successive values of qubit index $k$ is apparent in Figure \ref{fig:CompareDirectWalkShort}. Note also that for each pair of successive qubits $k$ and $k+1$, there are particular times for which the values of $C_k(t)$ and $C_{k+1}(t)$ are nearly identical. For example $C_2(t)$ and $C_3(t)$ are nearly the same for $t/\tau \approx 0.7$. Calculations on a fine grid of times around such points reveal that the values do not in fact become identical, merely quite close. 

The computational cost of using  direct operator exponentiation as in \refeq{LRexponentialDef} becomes prohibitive for chains of even 15 qubits. Moreover quantum reflections from the end of the chain propagate back and so relatively short chains can give a misleading impression of the way in which correlations would propagate in a longer chain. In the case shown in Figure  \ref{fig:CompareDirectWalkShort}, for example, after about $t/\tau =2$ the behaviour observed is strongly influenced by interference due to the finite chain length and reflections from the end. The method of operator Pauli walks gives us the ability to examine much larger systems and observe behavior far from the ends. We will explore this behavior more in Section \ref{sec:Propagation}.

\section{The semi-infinite chain at the phase transition \label{sec:CriticalPointCalculation}}

The results for $C_k(t)$ given by \refeq{LRcorrArb} can be simplified even further if we consider the case of a semi-infinite chain at the critical inter-qubit coupling $J'=1$.  The adjacency matrix in \refeq{AdjacencyArbNq} becomes
\begin{equation}
\bm{A}'_c =
\begin{bmatrix*}[r]
0      & 1     & 0      & 0      & 0       & \cdots \\
-1     & 0     & 1      & 0      & 0       & \cdots \\
 0     &-1     & 0      & 1      & 0       & \cdots \\
0      & 0     & -1     & 0      & 1       & \cdots \\
0      & 0     & 0      &-1      & 0       & \cdots \\
\vdots &\vdots &\vdots  & \vdots & \vdots  & \ddots
\end{bmatrix*}.
\label{eq:AdjacencyMatrixCritical}
\end{equation}
It will prove helpful in this context to renumber the operator nodes in \refeq{SigmaMLongChain} (not the qubits) so the first node has the index $0$. We then define
\begin{equation}
    G_{n,m} \equiv \left[(\bm{A}_c)^n \right]_{1,m+1},
\end{equation}
and write the correlation function
\begin{equation}
    C_k(t) = \sqrt{ \sum_{m=2k-1}^{2N_q-1} 
    \left| \sum_{n=0}^{\infty} \frac{2}{n!} \left(-2\pi \frac{t}{\tau} \right)^n G_{n,m}  \right|^2 }
    \label{eq:LRcorrCrit0}
\end{equation}
The quantity $G_{n,m}$  is the sum of the weight products for all walks of length $n$ from  node 0 to node $m$. The weight product for each of the walks will be $\pm 1$.
Each  negative step (a step to the left) in the walk generates a negative sign in the weight product.  The overall sign of the weight product for a walk is determined by the number of negative steps. All of the walks of length $n$ from node $0$ to node $m$ must have the same number of negative steps, and thus the same overall sign. One corollary is that in computing $G$ there is no interference between walks to $m$ that have the same length.

Consider a walk from node 0 to node $m$ of length $n$. Let $n_p$ be the number of positive steps and $n_n$ be the number of negative steps. The weight product for the walk will be $(+1)^{n_p}(-1)^{n_n}$. We have $n=n_p+n_n$ and $m=n_p-n_n$,
so
\begin{equation}
n_n=\frac{n-m}{2} \qquad \text{and}\qquad n_p=\frac{n+m}{2}.
\end{equation}

Since $n_n$ is an integer, we must have $n-m \equiv 0 \pmod{2}$, i.e., $n$  and $m$ must have the same parity. A walk with an even (odd) number of steps cannot end on an odd (even) node. 

 Let $N_w(n,m)$  be the \emph{number} of walks from node $0$ to node $m$ of length $n$. A well established result in combinatorics (Bertand's ballot problem) is that in the case of a line of nodes unbounded to the right the number of such walks is 
\begin{equation}
N_w(n,m) = 
\begin{cases}
\frac{m+1}{ 1+ (n+m)/2 } \binom{n}{(n+m)/2},\\ 
            \qquad\qquad \text{if }(n - m)=0 \pmod{2}  \nonumber\\
 0, \quad\qquad  \text{if }(n - m)=1 \pmod{2}  .
\end{cases}
\label{Walkcount1}
\end{equation}
(See, for example, the expression in reference \cite{RenaultBallotTheorem2008} with $a=n_p+1$ and $b=n_n$.) Note that  for $m > n, N_w(n,m)=0$.

Combining the information about the number of walks and the weight product for each walk, we can write
\begin{equation}
    G_{n,m} = (-1)^{(n-m)/2} \; N_w(n,m).
\end{equation}
So, if we consider a semi-infinite line of qubits ($N_q\rightarrow \infty$) we can write for the correlation function
\begin{widetext}
\begin{equation}
    C^2_k(t) = \sum_{m=2k-1}^{\infty} 
    \left| \sum_{n=0}^{\infty} \frac{2}{n!} \left(-2\pi \frac{t}{\tau} \right)^n (-1)^{(n-m)/2} \; N_w(n,m)  \right|^2 .
    \label{eq:LRcorrCrit}
\end{equation}
\end{widetext}

Because only values of $n$ with the same parity as $m$ contribute to the sum in \refeq{LRcorrCrit}, we  make the change of variables $n'\equiv (n-m)/2$, $n=2n'+m$ , with the result
\begin{multline}
    C^2_k(t) = 4 \sum_{m=2k-1}^{\infty} 
    \left(  \frac{2\pi t}{\tau}\right)^{2m}
    (m+1)^2 
        \\ \times \left| \sum_{n=0}^{\infty}
 \frac{(-1)^n}{(n+m+1)! n!} \left(\frac{2\pi t}{\tau} \right)^{2n} \right|^2 ,
    \label{eq:LRcorrCrit1}
\end{multline}
where we have then substituted the symbol $n$ for $n'$.
The sum over $n$ can be connected to a Bessel function  of the first kind through the relation (8.440 in \cite{GradshteynRyzhik1965})
\begin{equation}
    J_{m+1}(2x) = x^{m+1} \sum_{n=0}^{\infty}
    \frac{(-1)^n}{(n+m+1)! n!}\; x^{2n}.
    \label{eq:BesselFromSum}
\end{equation}
Substituting \refeq{BesselFromSum} into \refeq{LRcorrCrit1} 
yields
\begin{multline}
    C^2_k(t) = 4 \sum_{m=2k-1}^{\infty} 
    \left(  \frac{2\pi t}{\tau}\right)^{2m}
    (m+1)^2 
    \\ 
    \times \left| \sum_{n=0}^{\infty}        
    \left( \frac{2\pi t}{\tau} \right)^{-(m+1)}
     J_{m+1}\left(\frac{4\pi t}{\tau}  \right) \right|^2 
         \label{eq:LRcorrCrit2}
\end{multline}
\begin{align}
C^2_k(t)&=  \frac{1}{ \left(\frac{2\pi t}{\tau}\right)^2 } 
     \sum_{m=2k-1}^{\infty} 
      (m+1)^2  
      J_{m+1}^2\left(\frac{4\pi t}{\tau} \right) \nonumber\\
      &= \frac{1}{ \left(z/2\right)^2 } 
        \underbrace {\sum_{m=2k-1}^{\infty} (m+1)^2 J_{m+1}^2(z)}_{S},
    \label{eq:LRcorrCrit3}
\end{align}
where we let
\begin{equation}
    z\equiv \frac{4\pi t}{\tau}.
    \label{eq:Zdef}
\end{equation}
We divide the sum $S$ into two parts 
\begin{multline}
    S=\underbrace {\sum_{m=1}^{\infty} (m+1)^2 J_{m+1}^2(z)}_{S_1} \\
    \quad- \underbrace {\sum_{m=1}^{m=2k-2} (m+1)^2 J_{m+1}^2(z)}_{S_2}.
    \label{eq:Scrit0}
\end{multline}
The first term can be written
\begin{align}
    S_1 &= \sum_{m=1}^{\infty} m^2 J_{m}^2(z) - J_1^2(z).
       \label{eq:Scrit1}
\end{align}
Now we can use the relations (11.4.7 in \cite{AbramowitzStegun1964}) 
\begin{equation}
    J_m^2(x)=\frac{2}{\pi} \int_0^{\pi/2} J_{2m}(2x \cos{\theta}) d\theta
\end{equation}
and 
\begin{equation}
     \sum_{m=1}^\infty m^2 J_{2m}(u) =\frac{u^2}{8}
\end{equation}
(8.513.4 in \cite{GradshteynRyzhik1965}) to write \refeq{Scrit1} as
\begin{align}
    S_1&=\int_0^{\pi/2} \frac{2}{\pi} \sum_{m=1}^{\infty} m^2 J_{2m}(2z\cos{\theta}) d\theta - J_1^2(z)\nonumber\\
    &=\int_0^{\pi/2} \frac{z^2}{\pi}  \cos^2{\theta} d\theta - J_1^2(z)\nonumber\\
    &=\frac{1}{4} z^2 -J_1^2(z).
    \label{eq:Scrit2}
\end{align}
Substituting \refeq{Scrit2} and \refeq{Scrit0} into \refeq{LRcorrCrit3}, we obtain

\begin{align}
    C^2_k(t)&=\frac{
     \frac{1}{4} z^2 -J_1^2(z) - 
     \sum\limits_{m=1}^{2k-2} (m+1)^2 J^2_{m+1}(z) }
     { \left(z/2\right)^2 } \nonumber\\
     &= \frac{
       \frac{1}{4} z^2  - 
     \sum\limits_{m=1}^{2k-1} m^2 J^2_{m+1}(z) }
     { \left(z/2\right)^2 }.
\end{align}
Substituting in \refeq{Zdef}, we  arrive at a closed-form solution for the Lieb-Robinson correlation function for the $J'=1$ case,

\begin{equation}
    C_{k}(t) = \frac{ \sqrt{\frac{1}{4}\left(\frac{4 \pi t}{\tau}\right)^2 -  
    \sum\limits_{m=1}^{2k-1} m^2 J^2_{m}\left (\frac{4 \pi t}{\tau} \right) }}{\pi t/\tau} .
    \label{eq:CkSeminfiniteJp1}
\end{equation}

\section{The leading edge of correlations \label{sec:Leading edge}}


\begin{figure}[bt]
    \centering
    \includegraphics[width=\columnwidth]{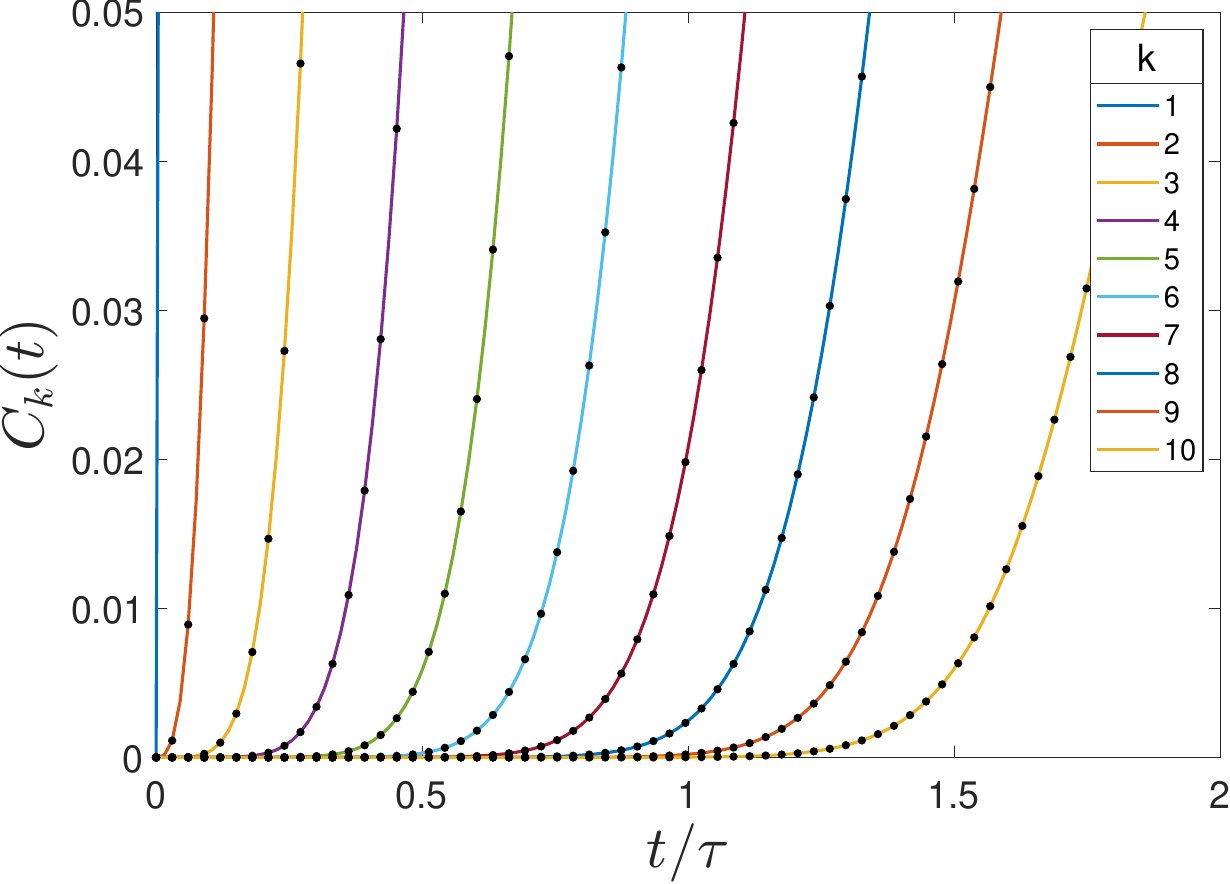}
    \caption{Leading edge of Lieb-Robinson correlations for a ten-qubit line with $J'=1/2$. Solid lines calculated directly from \refeq{LRexponentialDef} and black dots are the results of the Pauli walk method given by \refeq{LRcorrArb}. }
    \label{fig:EarlyTimeDirectWalkerLinear}
\end{figure}


The complicated behavior of the Lieb-Robinson correleation function shown in  Figure \ref{fig:CompareDirectWalkShort} is much simpler if we focus on the  the early turn-on of correlation in each successive bit---the leading edge of the correlation. Figure \ref{fig:EarlyTimeDirectWalkerLinear} shows $C_k(t)$ for $J'=1/2$ as correlations first start to grow and  propagate along the qubit chain. The solid lines show the  results computed directly from \refeq{LRexponentialDef} and the dots indicate the results from the Pauli walk method of \refeq{LRcorrArb}.   For each qubit $k$, as the correlation with qubit 1  just begins to grow, and $C_k(t)$ is still small, the growth follows a power-law behavior in time $\sim t^{2k-1}$.  At this leading edge, we have shown in  reference \cite{Mahoney2022}  that  the  Lieb-Robinson correlation function  is given  by:
\begin{equation}
C_k(t) \quad \underset{\text{\small leading edge}}{\xrightarrow{\hspace{1cm}}}
\quad \frac{{{2^{2k}}{\pi ^{2k - 1}}}}{{(2k - 1)!}}{\left( J'\right)^{k - 1}}{\left( {\frac{t}{\tau }} \right)^{2k - 1}}.
\label{eq:LR_EarlyAnalyticExact}
\end{equation}
Only odd orders contribute to the initial growth so the next most important term is proportional to $t^{2k+1}$.

\begin{figure}[tb]
    \centering
    \includegraphics[width=\columnwidth]{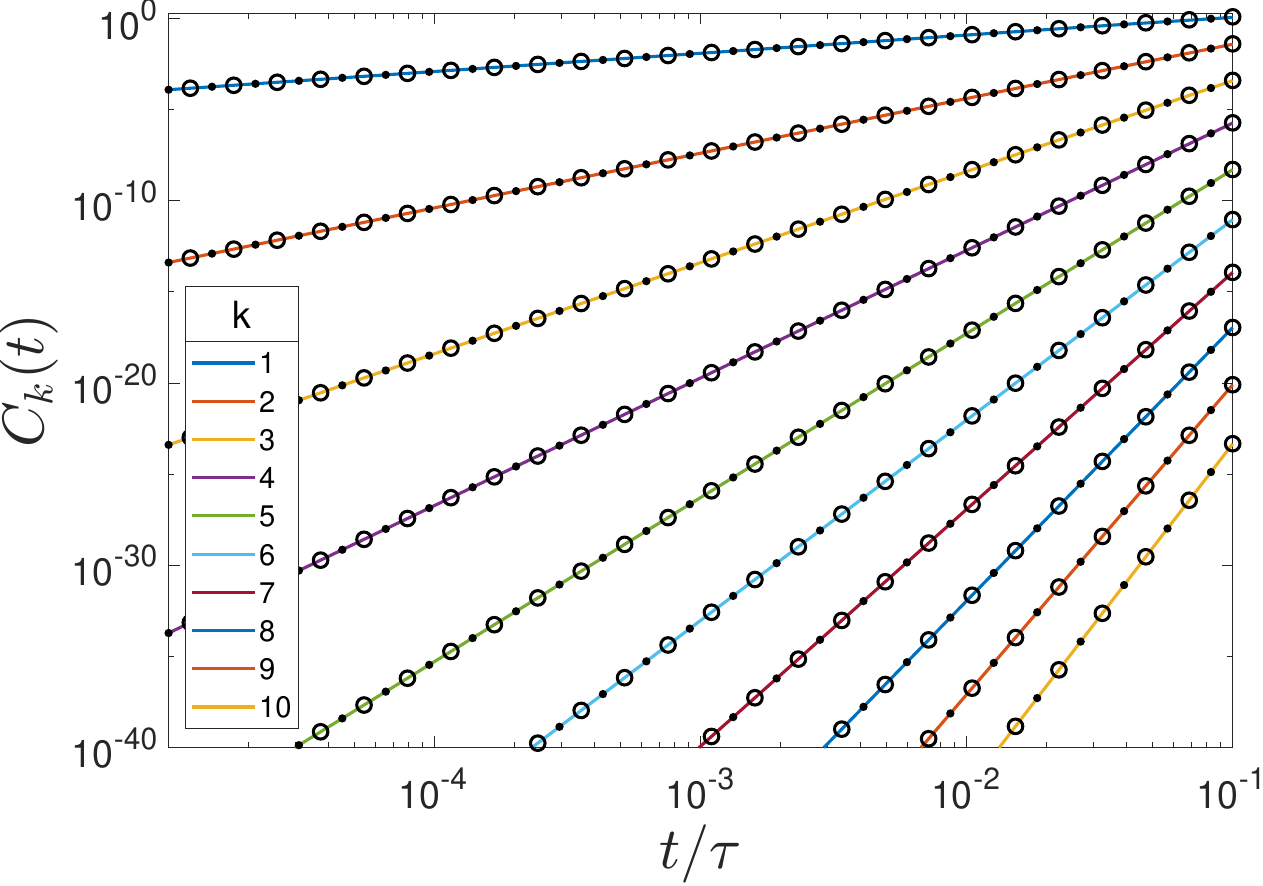}
    \caption{Leading edge of Lieb-Robinson correlations for a ten-qubit line with $J'=1/2$. The solid curves shows the results for $C_k(t)$ calculated using the exponential time evolution operator in \refeq{LRexponentialDef}. The black dots show the results calculated from the method of Pauli walks using \refeq{LRcorrArb}, and the open circles are calculated from the leading edge expression of \refeq{LR_EarlyAnalyticExact}.
    }
    \label{fig:EarlyTimeDirectWalkerLog}
\end{figure}

Figure \ref{fig:EarlyTimeDirectWalkerLog} shows the early growth of the Lieb-Robinson correlation function for the ten-qubit chain on a log-log scale. Three methods of calculating the correlation are shown: the direct exponential time dependence of \refeq{LRexponentialDef} (solid lines), the operator Pauli walk method of \refeq{LRcorrArb} (solid dots), and the leading-edge expression of \refeq{LR_EarlyAnalyticExact} (open circles). The agreement between the three methods is excellent. 


The leading edge correlation response quantifies the way in which the correlation between qubit $k$ and qubit $1$ initially turns on. How fast does this leading edge of correlation propagate down the chain?  For qubits far enough down the chain that we can replace the factorial in the denominator of   \refeq{LR_EarlyAnalyticExact} with Stirling's approximation, we obtain 
\begin{equation}
    C_k(t)\underset{ 
            \underset{\text{\tiny leading edge}} 
                     {\text{\tiny k large} }}
            {\xrightarrow{\hspace{1cm}}}
    \;\frac{1}{\sqrt{\pi J'}} 
    \frac{1}{\sqrt{k}}
    \left( \frac{v_{\text{\tiny LR}}t}{(k-1/2)} \ \right)^{2k-1},
    \label{eq:LR1DlargeKlimit}
\end{equation}
where the value of $v_{\text{\tiny LR}}$ is  given by
\begin{equation}
    v_{\text{\tiny LR}}\tau =e\pi \sqrt{J'} ,
    \label{eq:LR1Dvelocity}
\end{equation}
which we identify as the Lieb-Robinson velocity. 

\begin{figure}
    \centering
    \includegraphics[width=\columnwidth]{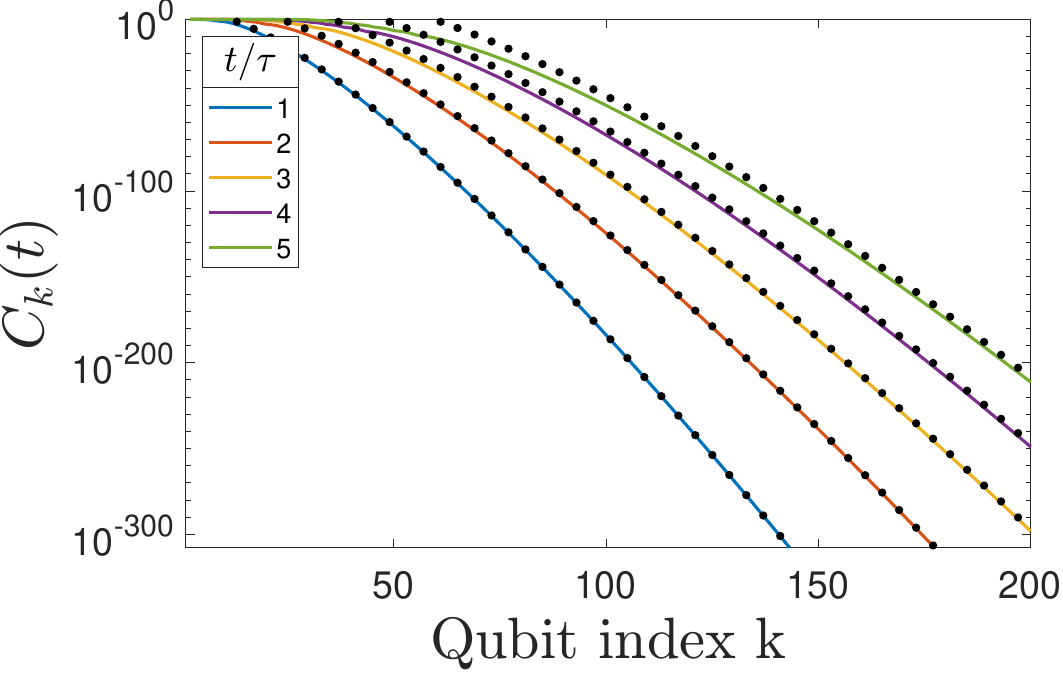}
    \caption{Leading edge of the Lieb-Robinson correlation function for $J'=2$. The solid lines show snapshots at particular times calculated with the full operator Pauli walk method of \refeq{LRcorrArb}. The black dots are calculated using the simplified  leading edge expression of \refeq{LR1DlargeKlimit}. 
    }
    \label{fig:LeadingEdgeCompariston}
\end{figure}

A comparison of the leading edge expression of \refeq{LR1Dvelocity} with the results of the operator Pauli walk method is shown in Figure \ref{fig:LeadingEdgeCompariston} for the $J'=2$ case. Snapshots of $C_k(t)$ are shown for various times as the correlations propagate down a chain of 200 qubits. The solid lines are from the Pauli walk method, which includes all higher orders in time. The calculation using the leading edge expression of \refeq{LR1Dvelocity} matches well  for early times, but not as well later as the correlations  moves down the chain and higher order terms become more important. As we will see in the next section, the correlation front moves down the chain and the leading edge description is accurate only well out ahead of that front.
Note that the magnitude of $C$ here is very small. 

To see something more like a simple exponential decay in front of the correlation front, one must go much further down the chain. In that limit, where $k$ is quite large and in the region of the leading edge, where $k\approx v_{LR}t$, \refeq{LR1DlargeKlimit} can be further simplified to
\begin{equation}
    C_k(t)
    \underset{ 
            \underset{\text{\tiny leading edge}} 
                     {\text{\tiny large k} }}
            {\xrightarrow{\hspace{1cm}}}
    \;\frac{e}{\sqrt{\pi J'}} 
    \frac{1}{\sqrt{k}}
    e^{ -2 \left(k-v_{\text{\tiny LR}} t \right) }.
    \label{eq:LR1DexponentialLimit}
\end{equation}
\begin{myhighlight}
The details of the connection between \refeq{LR1DlargeKlimit} and \refeq{LR1DexponentialLimit} are in Appendix A. 
\end{myhighlight}

\begin{figure}
    \centering
    \includegraphics[width=\columnwidth]{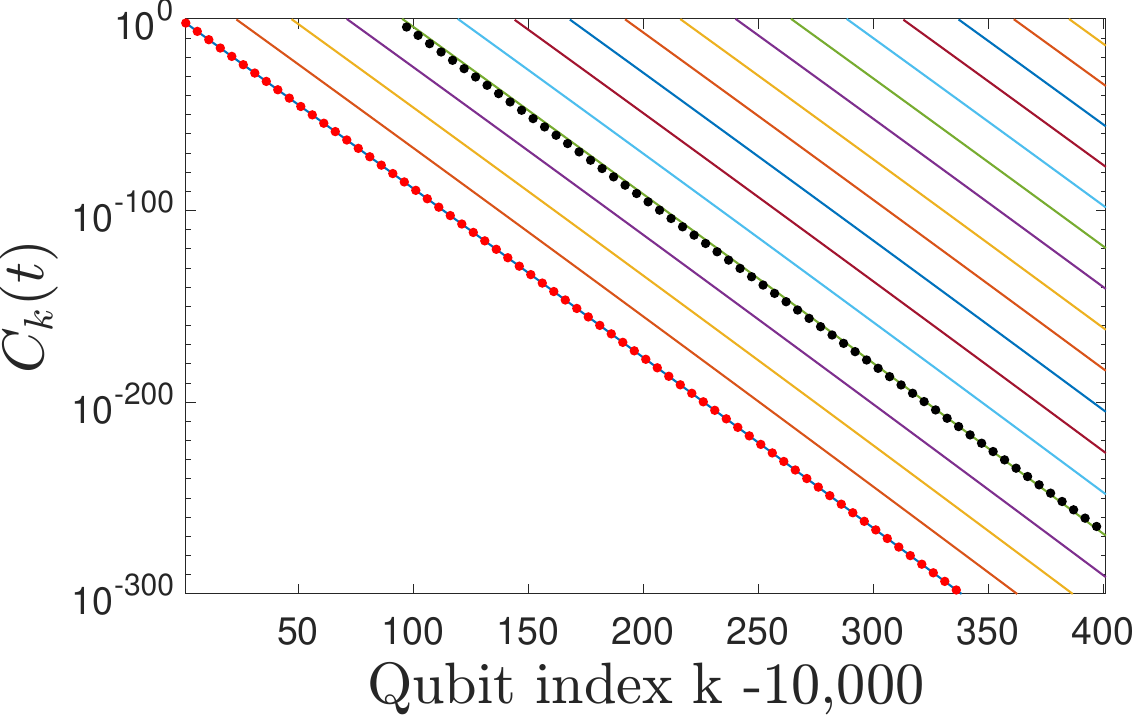}
    \caption{The leading edge of the Lieb-Robinson correlation function far down a qubit chain for $J'=2$. Snapshots of $C_k(t)$ are shown for even values of $t/\tau$ between 828 and 862. The solid lines are calculated using \refeq{LR_EarlyAnalyticExact}. Red dots are the results of  \refeq{LR1DlargeKlimit}, and black dots are calculated using \refeq{LR1DexponentialLimit}. 
    }
    \label{fig:LeadingEdgeFarLog}
\end{figure}

We recognize that this nearly-exponential leading edge of correlation corresponds to the classic results of Lieb and Robinson.   For the TFIM, we have here the additional information of the specific prefactor and its dependence on $J'$, and the  modification of the wavefront by the factor $1/\sqrt{k}$ (also noted in \cite{Chessa2019}). Figure \ref{fig:LeadingEdgeFarLog} shows a comparison between the results of \refeq{LR_EarlyAnalyticExact} (lines), \refeq{LR1Dvelocity} (red dots), and the exponential form given by \refeq{LR1DexponentialLimit} (black dots). For $C_k(t)$ to assume this simple form---an exponential front moving down the chain at velocity $v_{\text{\tiny LR}}$---requires looking very far down the line, here more than 10,000 qubits.

\section{Propagation of correlations \label{sec:Propagation} }

\begin{figure}[tb]
    \includegraphics[width=\columnwidth]{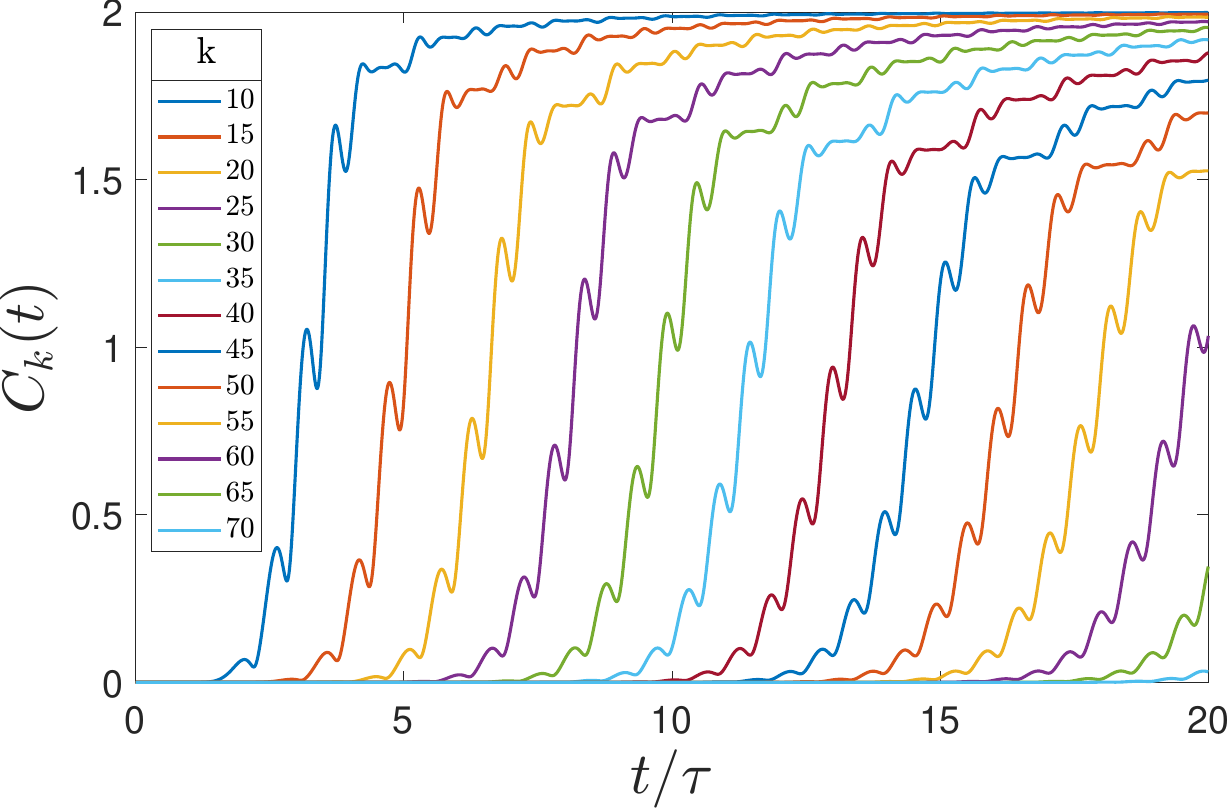}
    \caption{Time dependence of the Lieb-Robinson correlation function for a chain of $N_q=200$ qubits with coupling $J'=1/2$.
    }
    \label{fig:CNq200Jp0p5vsTime}
\end{figure}

\begin{figure}
    \centering
    \includegraphics[width=\columnwidth]{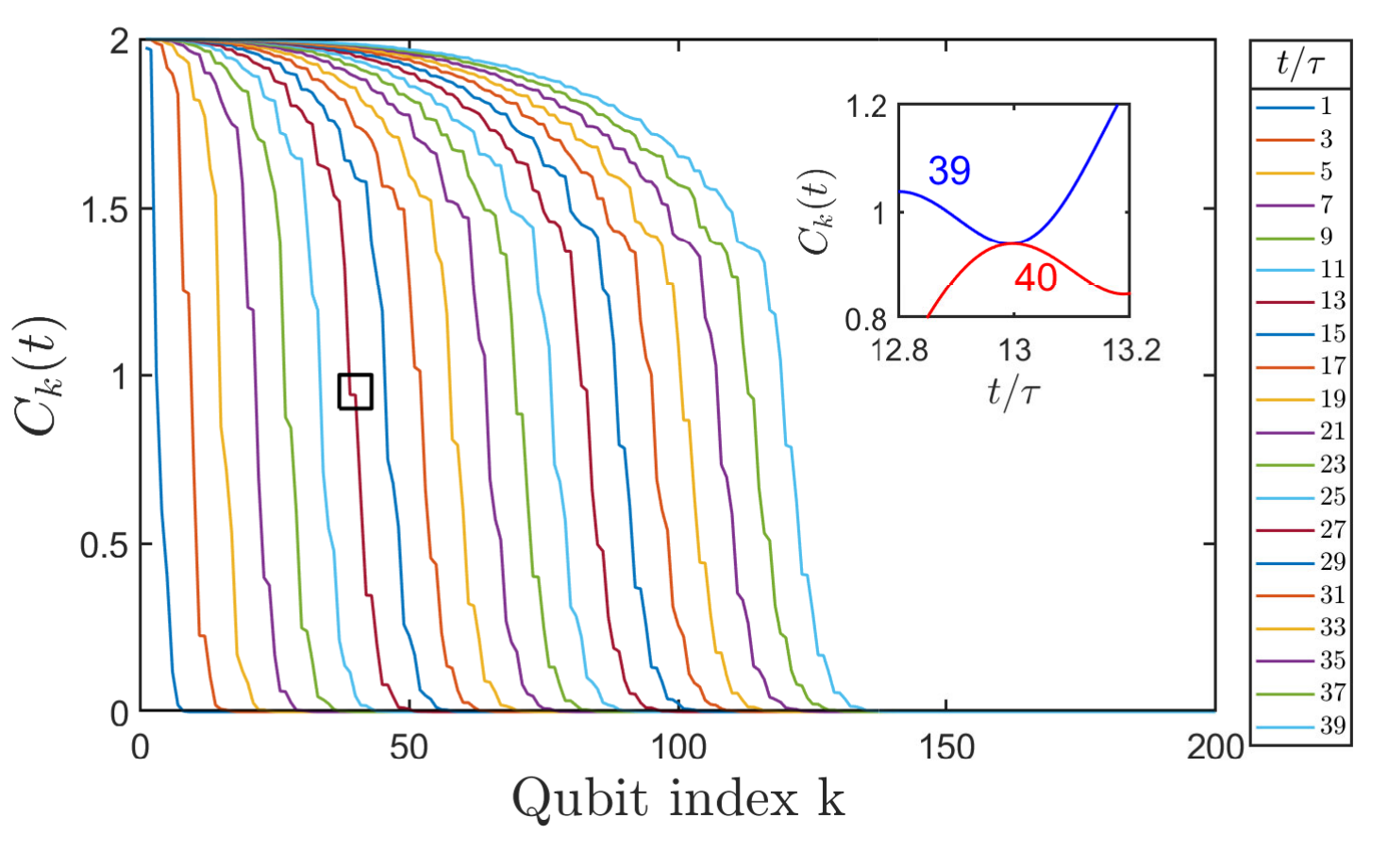}
    \caption{Snapshots of the Lieb-Robinson correlation function at different times for a chain of $N_q=200$ qubits with coupling $J'=1/2$. The box shows a plateau in $C_k(t)$ at  $t/\tau \approx 13$. The inset shows the time-dependence of $C_{39}$ and $C_{40}$ around that time. The near-equivalent values at these adjacent qubits produces the plateau.
    }
    \label{fig:CNq200Jp0p5vsK}
\end{figure}


We now have three techniques that let us characterize the way the correlation quantified by $C_k(t)$ propagates in larger systems. The operator Pauli walk method described in Section \ref{sec:PauliWalkMethod} is our primary tool, augmented by the critical point ($J'=1$) results from Section \ref{sec:CriticalPointCalculation}, and by the expression for the leading edge of correlations from Section \ref{sec:Leading edge}.

\subsection{Overall behavior and effect of coupling strength}

Figure \ref{fig:CNq200Jp0p5vsTime} shows the Lieb-Robinson correlation function calculated using the method of operator Pauli walks for a line with $N_q=200$ qubits and $J'=1/2$. The figure shows $C_k(t)$ as a function of time for qubit indices $k=[10, 15, 20, \ldots, 70]$ and times up to $20 \tau$. 

Several features of the way in which the correlation spreads down the chain now become clearer.  As one might expect, away from the ends, the turning on of correlations in each qubit becomes more regular, with a similar pattern for successive values of qubit index $k$ down the line. In the time-frame shown, reflections from the end of the chain play no role. The characteristic quantum oscillations remain, but now we can see that in the absence of end effects, $C_k(t)$ saturates to the maximum value of 2 eventually for all values of $k$. This feature was not as obvious in the shorter chain of Figure \ref{fig:CompareDirectWalkShort}.

While Figure \ref{fig:CNq200Jp0p5vsTime}  plots the time dependence of $C_k(t)$ for specific qubits down the chain,    Figure \ref{fig:CNq200Jp0p5vsK} shows the same process in a different way. The figure shows  $C_k(t)$ for all the qubits in the chain at particular snapshots in time: $t/\tau = [1, 3, 5, \ldots, 39]$. This displays the correlation front moving down the chain. Behind the correlation front, $C_k$ saturates at the maximum value of 2 (for this case when $J'=1/2$).   The regular spacing of the curves in both Figures \ref{fig:CNq200Jp0p5vsTime} and \ref{fig:CNq200Jp0p5vsK} suggests that the front moves with a constant velocity down the chain, creating  the so-called ``light cone'' of propagating influence. (It will turn out that this is not the Lieb-Robinson velocity.) At a time far in advance of the front's arrival, the state of qubit $k$ is necessarily independent of whatever is happening, or has happened, with qubit $1$ so $C_k(t)$ is small. The structure of the Hamiltonian has not yet allowed qubits 1 and $k$ the possibility of becoming quantum entangled with each other. 

We note that Figure \ref{fig:CNq200Jp0p5vsK} shows an initially perplexing feature---what appears to be small plateaus in $C_k$ as a function of $k$   at particular times. As the inset shows, this is simply a manifestation of the above-described near-equalities that occasionally occur for neighboring qubits. The plateau in the small box is simply due to the fact that  $C_{39}(t)$ and $C_{40}(t)$ are very nearly equal at $t/\tau \approx 13$.


\begin{figure}[tb]
    \centering
    \includegraphics[width=\columnwidth]{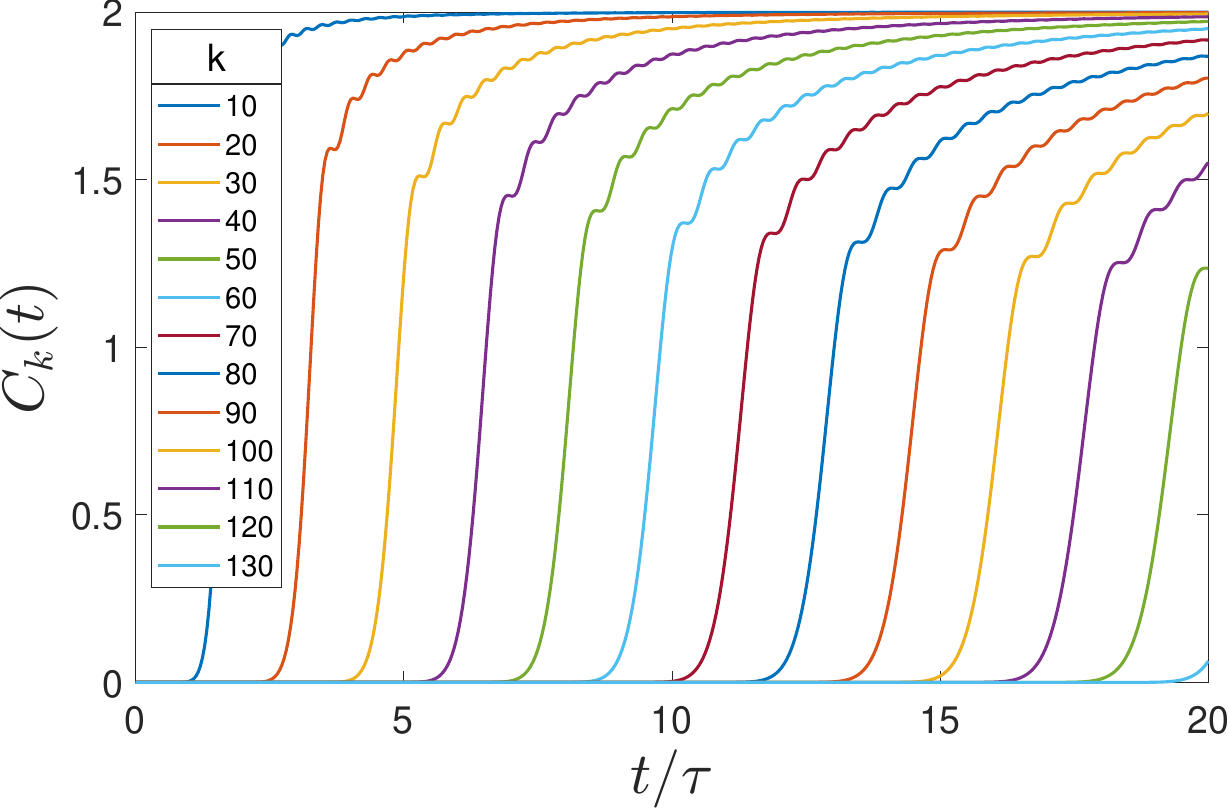}
    \caption{
    Time dependence of the Lieb-Robinson correlation function for a chain with coupling $J'=1$. This is the critical point for the quantum phase transition between the disordered ($J'< 1$) and ferromagnetic ($J'>1$) ground states.}
    \label{fig:CNq200Jp1p0vsTime}
\end{figure}

\begin{figure}[tb]
    \centering
    \includegraphics[width=\columnwidth]{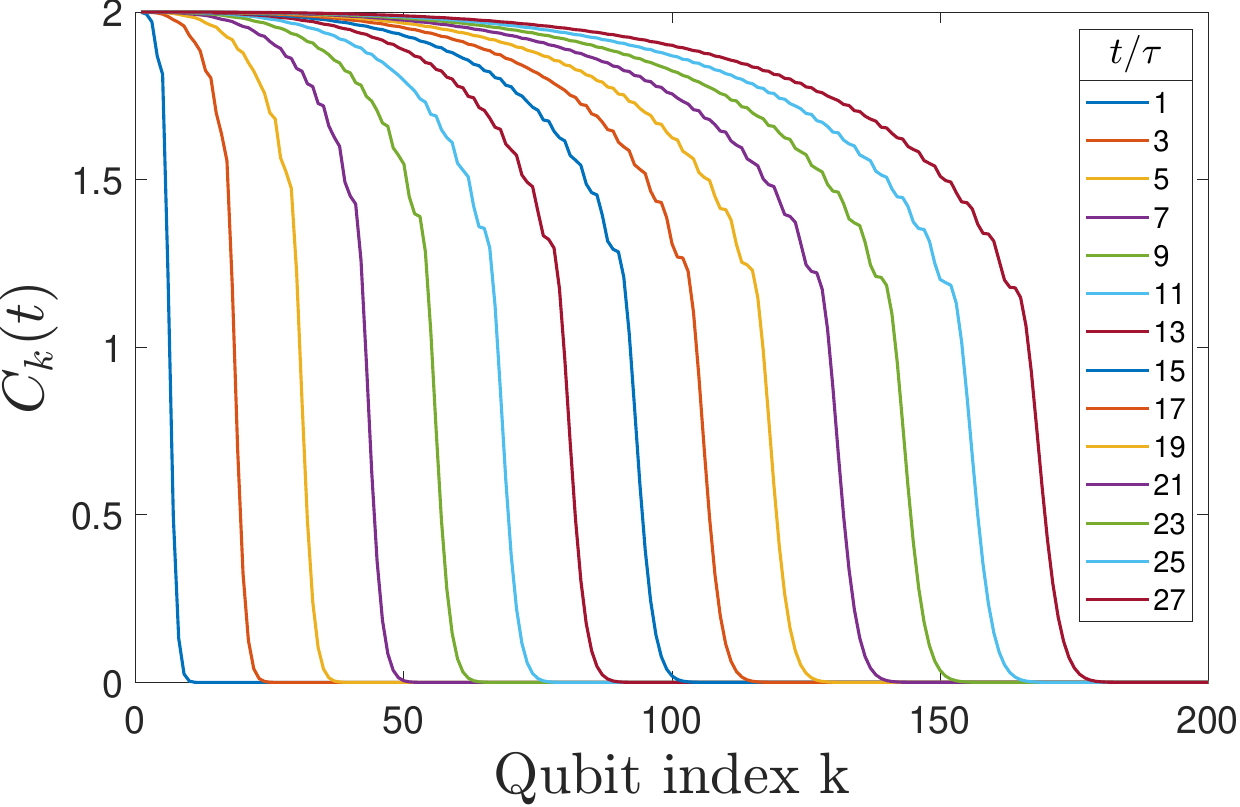}
    \caption{
    Snapshots of the Lieb-Robinson correlation function at different times for a chain of qubits with critical coupling $J'=1$. The correlation front is moving down the chain much faster than in the $J'=1/2$ case shown in Figure \ref{fig:CNq200Jp0p5vsK}.
    }
    \label{fig:CNq200Jp1p0vsK}
\end{figure}

Results for the propagation of correlations for the critical coupling $J'=1$ are shown in Figures \ref{fig:CNq200Jp1p0vsTime} and \ref{fig:CNq200Jp1p0vsK}.  In this case, we can use the analytic result for a semi-infinite chain of qubits given by \refeq{CkSeminfiniteJp1}. The behavior is qualitatively similar to the $J'=1/2$ case discussed above, although the curves are smoother. The values of $C_k(t)$ still saturate at $2$. There are no near-equalities for neighboring qubits and so no plateaus. One noteworthy difference that is immediately visible is that the correlation front moves down the chain much faster than the case with $J'=1/2$ (compare Figures \ref{fig:CNq200Jp0p5vsK} and  \ref{fig:CNq200Jp1p0vsK}). We will return to analysis of the velocity of propagation below.


\begin{figure}[tb]
    \centering
    \includegraphics[width=\columnwidth]{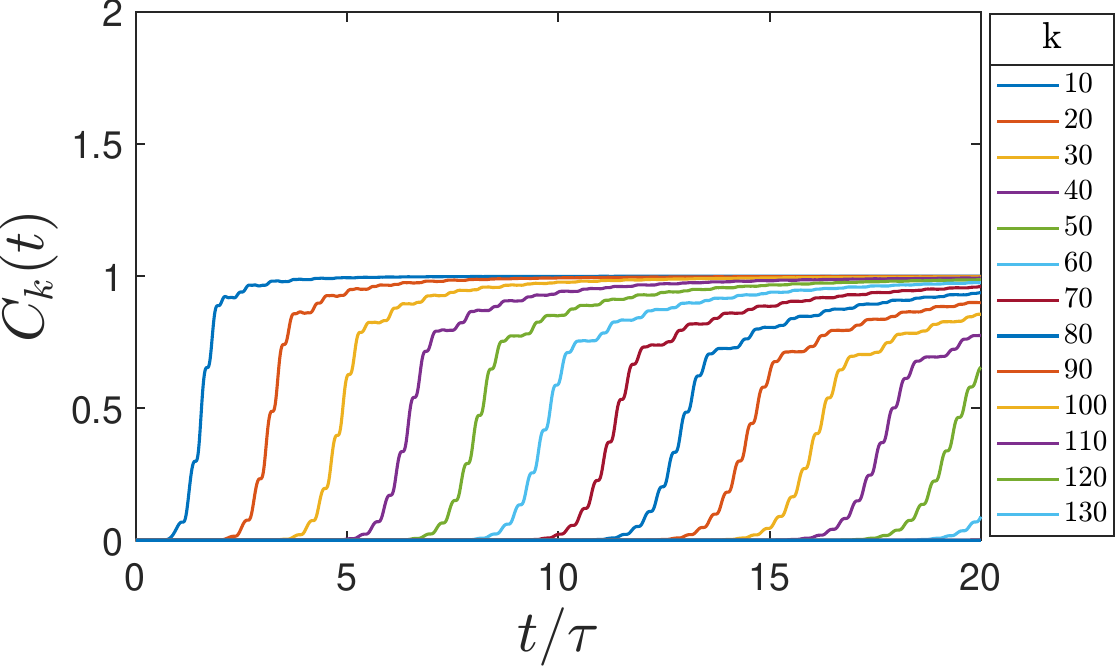}
    \caption{
    Time dependence of the Lieb-Robinson correlation function for a chain of $N_q=200$ qubits with coupling $J'=2$. 
    }
    \label{fig:CNq200Jp2p0vsTime}
\end{figure}

\begin{figure}[tb]
    \centering
    \includegraphics[width=\columnwidth]{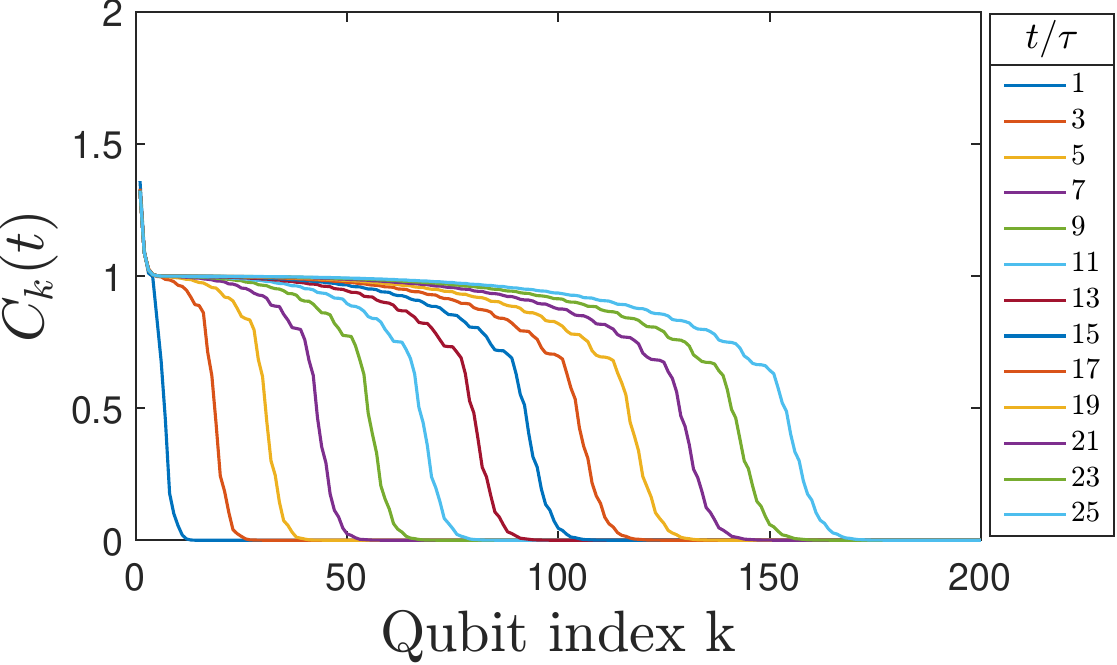}
    \caption{
    Snapshots of the Lieb-Robinson correlation function at different times for a chain of $N_q=200$ qubits with  coupling $J'=2$.
    }
    \label{fig:CNq200Jp2p0vsK}
\end{figure}

The correlation function in the case of stronger coupling, here $J'=2$, is shown in Figures \ref{fig:CNq200Jp2p0vsTime} and \ref{fig:CNq200Jp2p0vsK}, for $N_q=200$ qubits. This stronger coupling produces a ferromagnetically ordered ground state with a double degeneracy.  These results are calculated with the method of Pauli walks.  Unlike either the weak-coupling or critical coupling case, the correlation function now saturates at less than the maximal value of $2$--in this case $C_k(t)$ saturates at the value $1$. The velocity of the correlation front is the same as that in the $J'=1$ case. 


\subsection{Saturation value}
For a long chain, where reflection from the end can be neglected, we see that $C_k(t)$ for any qubit eventually reaches a saturation value which we denote $C_{\text{sat}}$. 
When the correlation front that is moving down the line is well passed a particular qubit $k$, the correlation function $C_k$ for that qubit will approach arbitrarily closely to $C_{\text{sat}}$.

\begin{figure}[tb]
    \centering
    \includegraphics[width=\columnwidth]{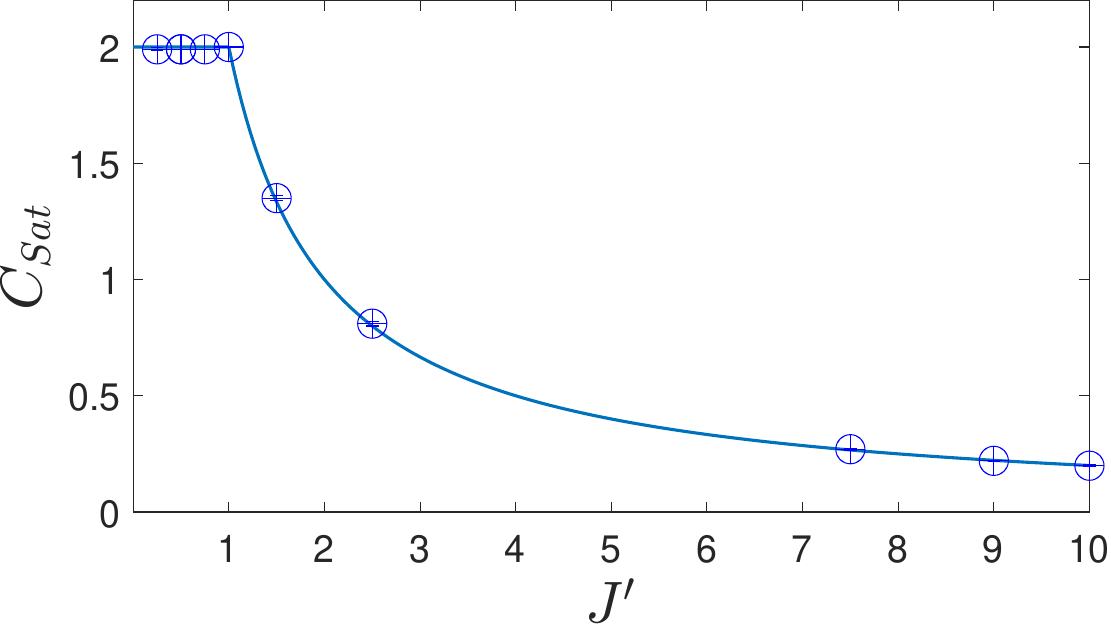}
    \caption{Saturation values of the Lieb-Robinson correlation function.}
    \label{fig:CsatVsJp}
\end{figure}

The calculated values of $C_{\text{sat}}$ for different values of $J'$ are shown in Figure \ref{fig:CsatVsJp}. For  $J'\le 1$, corresponding to the weakly-coupled disordered  phase, $C_{\text{sat}}=2$ independent of $J'$.  We observe that for  $J'\geq 1$, corresponding to the ordered phase, the value of $C_{\text{sat}}$ decreases inversely with increasing $J'$: 
\begin{align}
    C_{\text{sat}}=
    \begin{cases}
        2, &\quad    J'\le 1 \\
        2/J', &\quad  J' > 1.
    \end{cases}
\end{align}
The agreement between the numerical calculations using the Pauli walk method  and this conjectured, though unproven, analytical expression is excellent.

\subsection{The velocity of correlation propagation}
We have seen in Figures \ref{fig:CNq200Jp0p5vsTime}--\ref{fig:CNq200Jp2p0vsK} that the Lieb-Robinson correlation front propagates down the chain of qubits. How fast does this front travel? We  quantify this in the obvious way by picking an arbitrary threshold level $C_{\text{threshold}}$ and finding the time $t_k$ when $C_k(t) = C_{\text{threshold}}$. Using finite differences, the velocity of the front can then be calculated. Near the beginning of the chain, the velocity  varies, but subsequently saturates to a stable value, as is clear in the figures. We find that this value $v_{\text{\tiny front}}$ is 
\begin{align}
    v_{\text{\tiny front}}\tau=
    \begin{cases}
       2\pi J'  &\quad    J'\le 1 \\
        2\pi         &\quad    J' > 1.
    \end{cases}
    \label{eq:vFrontResults}
\end{align}


This front velocity can be connected to a well-established result for the QTFIM. For the infinite qubit chain, or one with periodic boundary conditions, the  Hamilonian of \refeq{IsingHamiltonian} can be mapped onto an equivalent fermion problem using the Jordan-Wigner transformation  \cite{LiebMattis1961, SachdevBook2011, Toth2023}. A subsequent Bogoliubov transformation rotates the fermion creation and annihilation operators to yield a free (non-interacting) fermion Hamiltonian with quasiparticle excitation spectrum 

\begin{equation}
 E(q)= 2J\sqrt{g^2 + 1 - 2g\cos{q}} .
 \label{eq:epsQuasiparticle}
\end{equation}
Here $q$ is the wavenumber of the harmonic excitation and $g\equiv 1/J'$. 
The corresponding group velocity can be calculated via the usual expression
\begin{equation}
 v_g(q)=\frac{1}{\hbar} \frac{\partial }{\partial q} E(q). 
 \label{eq:vGroup}
\end{equation}

\noindent The maximum value of this group velocity $v^{\text{\tiny max}}_g$ is given by precisely the expression for $v_{\text{\tiny front}}$ in \refeq{vFrontResults}. The derivation is included for completeness in 
\begin{myhighlight}
    Appendix B.
\end{myhighlight}  Thus, the correlation front quantified by $C_k(t)$ and depicted in Figures \ref{fig:CNq200Jp0p5vsTime}--\ref{fig:CNq200Jp2p0vsK} travels down the qubit chain at the speed of the maximum quasiparticle group velocity. 

\begin{figure}[t!]
    \centering
    \includegraphics[width=\columnwidth]{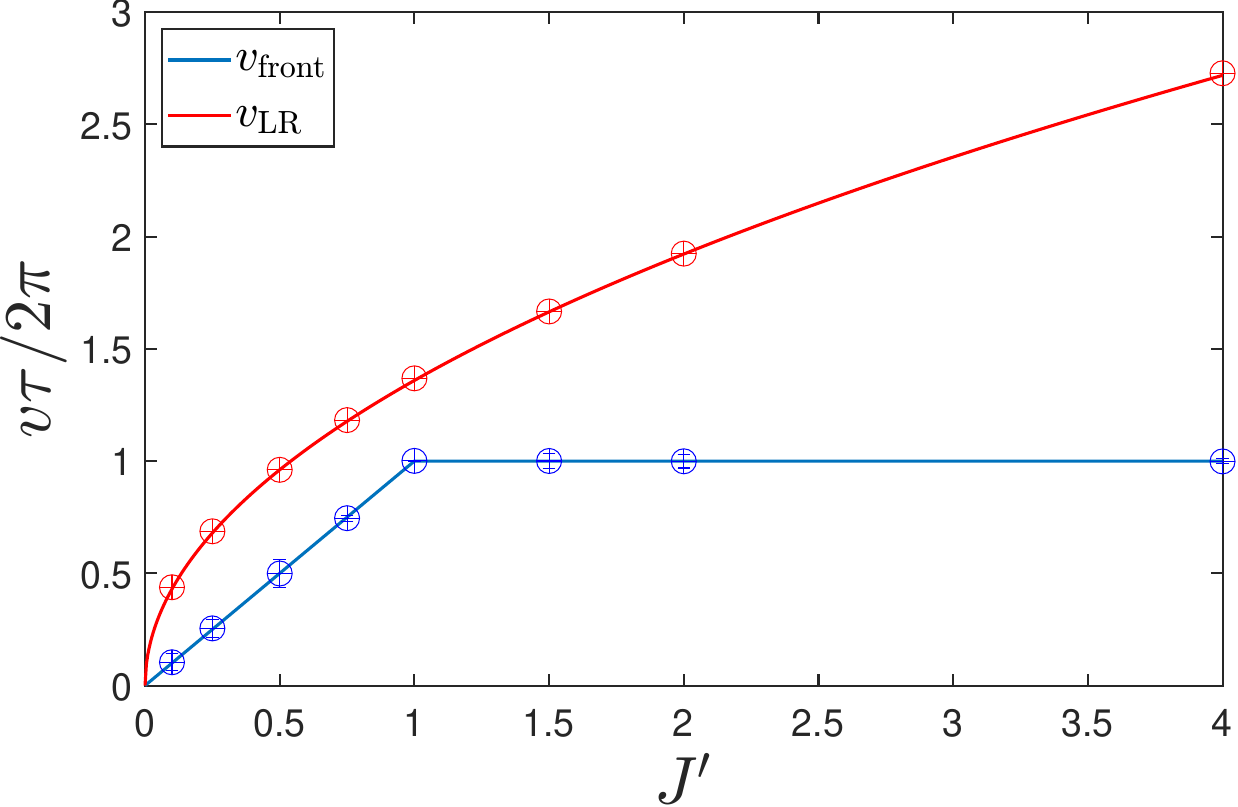}
    \caption{Correlation front velocity and the Lieb-Robinson velocity for the leading edge. Points are from numerical solutions using the operator Pauli walk method. Solid lines are from Eqs. (\ref{eq:LR1Dvelocity}) and (\ref{eq:vFrontResults}). } 
    \label{fig:SaturationVelocities}
\end{figure}

There is another velocity in the problem, as we have already described in Section \ref{sec:Leading edge}, the Lieb-Robinson velocity in Eqs. (\ref{eq:LR1DlargeKlimit}) and (\ref{eq:LR1Dvelocity}). The front velocity $v_{\text{\tiny front}}$ and the Lieb-Robinson velocity $v_{\text{\tiny LR}}$ are plotted as a function of coupling strength $J'$ in Figure \ref{fig:SaturationVelocities}. The points are obtained from numerical calculations using the operator Pauli walk method. The solid lines are from the analytic expressions in Eqs. (\ref{eq:LR1Dvelocity}) and (\ref{eq:vFrontResults}). The phase transition at $J'=1$ is clearly evident in $v_{\text{\tiny front}}$, but is not manifested in $v_{\text{\tiny LR}}$.
For any value of $J'>0$,  
\begin{equation}
v_{\text{\tiny LR}} > v_{\text{\tiny front}},
\end{equation}
so the exponential leading edge of correlation is traveling down the chain considerably ahead of, and faster than, the main correlation front seen in Figures \ref{fig:CNq200Jp0p5vsTime}--\ref{fig:CNq200Jp2p0vsK}. 

\begin{figure}[bt]
    \centering
    \includegraphics[width=\columnwidth]{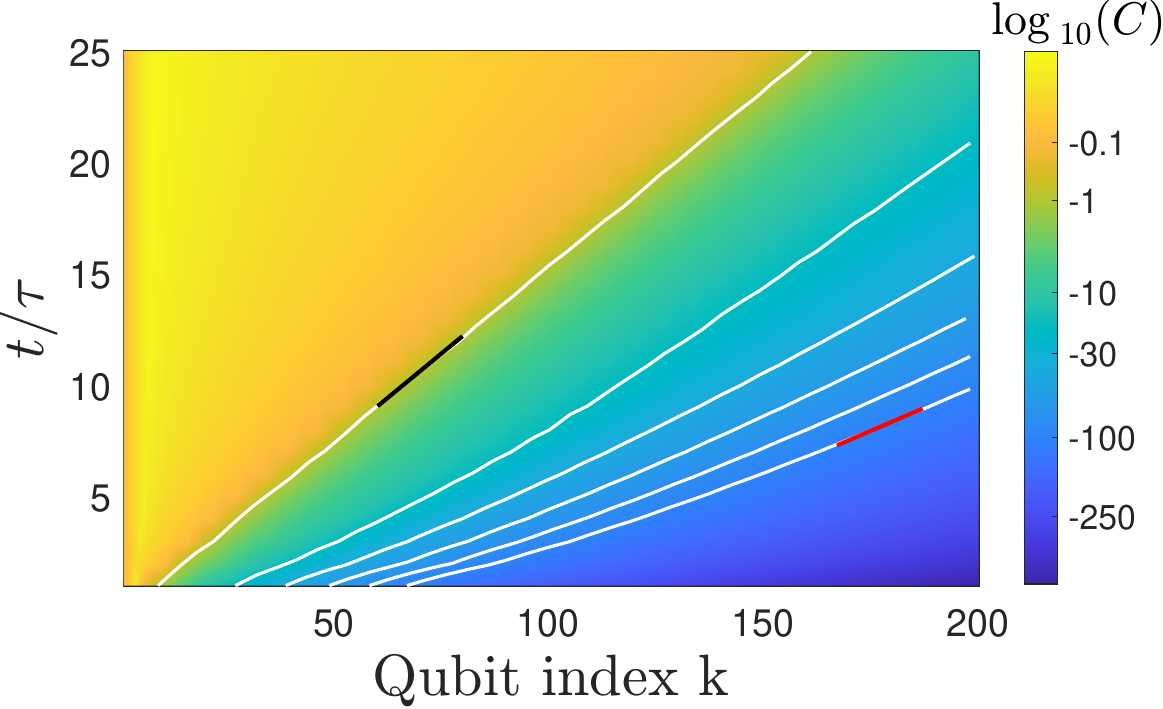}
    \caption{
    The ``light-cone'' plot of the Lieb-Robinson correlation function for a 200 qubit chain with $J'=2$.  Isocontours are shown for 
    $\log_{10}(C_k(t))=[-1, -20, -40, -60, -80, -100]$. The slope of the solid black line segment corresponds to $v_{\text{\tiny front}}$ given by \refeq{vFrontResults}. The slope of the red line segment corresponds to $v_{\text{\tiny LR}}$ given by \refeq{LR1Dvelocity}.
    }
    \label{fig:LightConeJp2}
\end{figure}

The relationship of the two velocities in the problem can also be seen in the light-cone plot of $C_k(t)$ in Figure \ref{fig:LightConeJp2}. The color in the figure represents the logarithm of the correlation function  for a 200-qubit chain with $J'=2$. The calculation is done using the Pauli walk method of \refeq{LRcorrArb}.  Isocontours for the correlation function are shown in white (slight oscillations are an artifact of the contouring algorithm and the discreteness of the grid). The black line segment corresponds to the velocity $v_{\text{\tiny front}}$, and the red line segment corresponds to $v_{\text{\tiny LR}}$.
Because time is on the vertical axis, a lower slope on the graph corresponds to a higher velocity.
The main correlation front expands with a linear light cone given by $v_{\text{\tiny front}}$. The velocity of the leading edge is initially faster than that given by \refeq{LR1Dvelocity}, and saturates at the Lieb-Robinson velocity from above \cite{Mahoney2022}. As the primary front expands down the chain, the slope of the isocontours at a particular qubit shift to correspond to $v_{\text{\tiny front}}$. The region that is sufficiently far enough ahead of the primary front that it is well-described by $C_k \sim e^{-2(k-vt)}$ is more than 100 qubits down the chain and has an extremely small value of the correlation function.   

\begin{myhighlight} 

A note about terminology is perhaps in order here. The original Lieb-Robinson paper \cite{LiebRobinson1972} used the term ``group velocity'' to describe the velocity that appears in \refeq{BasicLRresult}. Their result was very general and did not depend on a specific energy dispersion relationship, or indeed on the precise Hamiltonian. The meaning of this velocity is provided by the context in which it appeared---a bound on correlations that depends on the quantity $(d - vt)$. Here we use the term  ``group velocity'' in the standard sense provided by \refeq{vGroup}. In this case, the group velocity $v_g$ is understood as the speed of a wavepacket composed of a superposition of single-quasiparticle Hamiltonian eigenstates with some distribution of wavenumbers centered about $q$. We identify our observed (calculated) front velocity $v_{\text{\tiny front}}$ with the maximum group velocity so defined. 

\end{myhighlight} 

\section{Discussion \label{sec:Discussion}}

The primary result of this investigation is to highlight the fact that for the 1D QTFIM two different velocities characterize the propagation of the Lieb-Robinson correlation function. A correlation front  moves  down the qubit chain with velocity $v_{\text{\tiny front}}$, ahead of which $C_k(t)$ is  exponentially smaller. This velocity increases linearly with coupling strength in  the disordered Ising phase ($J'<1$) and is independent of coupling strength in the ordered phase. The velocity is given by the maximum group velocity of the single quasiparticle excitation band. Behind the propagating front, $C_k(t)$ saturates to a value of 2 in the disordered phase, and $2/J'$ in the ordered phase. 

Well ahead of the correlation front, the exponential leading edge of the correlation front moves faster. This velocity saturates from above to the Lieb-Robinson velocity $v_{\text{\tiny LR}}=e \sqrt{J\gamma}/\hbar$ (equivalent to \refeq{LR1Dvelocity}). This velocity is unaffected by the phase transition at $J'=1$. The Lieb-Robinson velocity describes a region well out ahead of the main correlation front in which the value of $C_k(t)$ is extremely small.


It is not clear that the operator Pauli walk method developed here for the QTFIM can be extended to other models. The key feature that enabled this method to work was that the number of Pauli strings that were required to support the spreading Heisenberg operator was reduced from $4^{N_q}$ to $2N_q$. For longer-range interactions, that may well not be the case. Further investigation will be required to reveal whether the general features of the propagation of $C_k(t)$ as described here hold for such systems.


\appendix 
\begin{myhighlight} 
\section{Derivation of Equation \refeq{LR1DexponentialLimit}}
We begin with  \refeq{LR1DlargeKlimit} and consider the limit in which the qubit index $k$ is quite large and focus on the region ahead of the leading edge of correlations defined by $k\approx v_{\text{\tiny LR}}t$. Equation~(\ref{eq:LR1DlargeKlimit}) can be written
\begin{equation}
    C_k(t)\underset{ 
            \underset{\text{\tiny leading edge}} 
                     {\text{\tiny k large} }}
            {\xrightarrow{\hspace{1cm}}}
    \;\frac{1}{\sqrt{\pi J'}} 
    \frac{1}{\sqrt{k}}
    \left[v_{\text{\tiny LR}}t \right]^{2k-1}\left[ \frac{1}{(k-1/2)} \ \right]^{2k-1}.
    \label{eq:LR1DlargeKlimit1}
\end{equation}

\begin{eqnarray}
    \left[ \frac{1}{k-\frac{1}{2}} \right]^{2k-1} &=& 
    \left[ \frac{1}{k(1-\frac{1}{2k}) } \right]^{2k-1} \nonumber\\
    &=&\left( \frac{1}{k}\right)^{2k-1} 
       \left[ \frac{1}{(1-\frac{1}{2k}) }\right]^{2k-1}\nonumber\\
     &=&\left( \frac{1}{k}\right)^{2k-1} 
       \left[ \frac{1}{(1-\frac{1}{2k}) }\right]^{-1}
       \left[ \frac{1}{\left(1-\frac{1}{2k}\right) }\right]^{2k}\nonumber\\
     &=&\left( \frac{1}{k}\right)^{2k-1} 
       \left[1-\frac{1}{2k} \right]
       \left[ \frac{1}{ \left( 1-\frac{1}{2k}) \right)} \right]^{2k} 
       \label{eq:LR1DlargeKPiece}
\end{eqnarray}
\noindent In the large $k$ limit, the first term in square brackets in \refeq{LR1DlargeKPiece} becomes 1. For the second term in square brackets, we can use the fact that 
\begin{eqnarray}
    \lim_{x\to\infty} \left[ \frac{1}{(1-\frac{1}{x}) }\right]^{x}=e
\end{eqnarray}
\noindent to obtain
\begin{equation}
    C_k(t)\underset{ 
            \underset{\text{\tiny leading edge}} 
                     {\text{\tiny k large} }}
            {\xrightarrow{\hspace{1cm}}}
    \;\frac{e}{\sqrt{\pi J'}} 
    \frac{1}{\sqrt{k}}
    \left[ \frac{v_{\text{\tiny LR}}t}{k} \right]^{2k-1}.
    \label{eq:LR1DlargeKlimit2}
\end{equation}

The term in square brackets in \refeq{LR1DlargeKlimit2} can be written
\begin{equation}
    \left[ \frac{v_{\text{\tiny LR}}t}{k}  \right]^{2k-1} = 
    e^{ (2k-1)  \log{ \left(  \frac{v_{\text{\tiny LR}}t}{k}   \right)} }.
    \label{eq:LR1DlargeKlimitExp1}
\end{equation}
We define the index at the leading edge of correlation at time $t$ to be 
\begin{equation}
k_t\equiv v_{\text{\tiny LR}}t 
\end{equation}
and consider a region of width $\Delta k$ out in front of this point so 
\begin{eqnarray}
k&=&k_t + \Delta k   \label{eq:DeltakDef1} \\
\Delta k &=&  k-k_t =k- v_{\text{\tiny LR}}t. \label{eq:DeltakDef2} 
\end{eqnarray}
We are interested in the limit when $k_t \rightarrow \infty$ and therefore $\Delta k/k_t\rightarrow 0$. In that case
\begin{equation}
\frac{v_{\text{\tiny LR}}t}{k} = \frac{k_t}{k_t + \Delta k} 
   = \frac{1}{1+\frac{\Delta k}{k_t}}
\end{equation}
so the logarithm in \refeq{LR1DlargeKlimitExp1} can be written
\begin{eqnarray}
\log{ \left(  \frac{v_{\text{\tiny LR}}t}{k}   \right)} 
&=&    -\log{ \left( 1 + \frac{\Delta k}{k_t}\right)  }\\
&\approx& -\frac{\Delta k}{k_t}.
\end{eqnarray}

The argument of the exponential in \refeq{LR1DlargeKlimitExp1} can therefore be written
\begin{eqnarray}
    (2k-1)  \log{ \left(  \frac{v_{\text{\tiny LR}}t}{k}   \right)}
    &\approx& -2k\frac{\Delta k}{k_t} + \frac{\Delta k}{k_t}.
\end{eqnarray}
The second term on the right can be dropped to first order, but the first term must be treated more carefully. Using \refeq{DeltakDef1}, we have
\begin{eqnarray}
(2k-1)  \log{ \left(  \frac{v_{\text{\tiny LR}}t}{k}   \right)}
   &\approx& -2(k_t + \Delta k) \frac{\Delta k}{k_t} \nonumber\\
    &\approx& -2\Delta k \nonumber\\ 
    &\approx& -2(k-v_{\text{\tiny LR}}t).
\end{eqnarray}

Combining this result with \refeq{LR1DlargeKlimitExp1} and \refeq{LR1DlargeKlimit2} we obtain
\begin{equation}
    C_k(t)
    \underset{ 
            \underset{\text{\tiny leading edge}} 
                     {\text{\tiny large k} }}
            {\xrightarrow{\hspace{1cm}}}
    \;\frac{e}{\sqrt{\pi J'}} 
    \frac{1}{\sqrt{k}}
    e^{ -2 \left(k-v_{\text{\tiny LR}} t \right) }
    \label{eq:LR1DexponentialLimitLast}
\end{equation}
which is identical to \refeq{LR1DexponentialLimit}. The validity of this expression is confirmed by the close match shown in Figure 6. Of course, if the width of $\Delta k$ is extended too far, then $C_k(t)$ will drop faster. The exponential given by \refeq{LR1DexponentialLimitLast} bounds this very advanced leading edge of the correlation function in a region where its magnitude is extremely small.

\end{myhighlight} 


\section{Maximum quasi-particle group velocity for the transverse Ising chain}

We start with the dispersion relation for single quasiparticle excitations for the QTFIM,
\begin{equation}
E(q)= 2J\sqrt{g^2 + 1 - 2g\cos{q}} \label{eq:initial},
\end{equation}
where $g=1/J'=\gamma/J$ \cite{SachdevBook2011}.

The group velocity is given by
\begin{equation}
v_g=\frac{1}{\hbar} \frac{d E(q)}{dq},
\end{equation}
and
\begin{align}
    \frac{d E(q)}{dq} &= \frac{2Jg \sin(q)}
    {\sqrt{ (g-\cos(q))^2 + \sin^2(q) }}.
    \label{eq:dEdkIsing2}
\end{align}

\noindent We consider two cases:

\medskip

\noindent {\bf Case I} \hspace{0.25in} $g\le 1, \qquad J'\ge 1$

\medskip

\noindent  If $g\le 1$, (\ref{eq:dEdkIsing2}) can be maximize by making
$(g-\cos(q)$ in the denominator $0$, so $q=\arccos(g)$.
Hence, 
\begin{align}
v_g^{\text{\tiny max}} &= \frac{1}{\hbar} \left(   
\frac{d E(q)}{dq} \right)_{\text{max}}= \frac{2Jg}{\hbar} \\
&=\frac{2 \pi}{\tau}.
\end{align}

Therefore
\begin{align}
v_g^{\text{\tiny max}}\tau = 2\pi \quad\quad \text{for }\; g \le 1, \quad J' \ge 1.
\end{align}

\medskip

\noindent  {\bf Case II } \hspace{0.25in} $g\ge 1, \qquad J'\le 1$

\medskip

We write \refeq{dEdkIsing2} as:
\begin{align}
\frac{d E_(q)}{dq} &  = \frac{2Jg}{\sqrt{1+  \left( \frac{g}{\sin(q)} - \cot(q) \right)  }^2}
\label{eq:deriv3}
\end{align}

\noindent Maximizing  (\ref{eq:deriv3}) is equivalent to minimizing 
\begin{equation} 
\beta(q)=\frac{g}{\sin{q}} - \cot{q}.
\end{equation}

\noindent Taking the derivative of $\beta$ with respect to $k$ and setting it to zero, we obtain:
\begin{align}
\frac{d \beta}{dq}  &= -g\cot(q)\csc(q) + \csc^2(q) =0,
\end{align}
so at the value $q_0$ which minimizes $\beta$
\begin{align}
g\cot(q_0)  =\csc(q_0), \quad
\cos(q_0)=\frac{1}{g}, \label{eq:cosk0}
\end{align}
and
\begin{align}
\sin(q_0) =\sqrt{1-\frac{1}{g^2}}.
\label{eq:sink0}
\end{align}

\noindent Substituting (\ref{eq:cosk0}) and (\ref{eq:sink0}) into (\ref{eq:dEdkIsing2}) we obtain
\begin{align}
    \left . \frac{d E(q)}{dq}\right|_{q_0}   
    =2J.
\end{align}
The maximum group velocity in this case is then
\begin{align}
    v_g^{\text{\tiny max}}     &= \frac{2J}{\hbar} 
\end{align}
or
\begin{align}
    v_g^{\text{\tiny max}}\tau &=  2\pi J'.
\end{align}

\noindent Summarizing the two cases therefore we have,
\begin{equation}
v_g^{\text{\tiny max}}\tau =
\begin{cases}
			2\pi J',   & J'\le 1, \quad g \ge 1\\
            2\pi,      & J' \ge 1 , \quad g \le 1.
\end{cases}
\label{eq:VgIsingFinal}
\end{equation}

%
%

%

\end{document}